\documentclass{article}

\usepackage{arxiv}

\usepackage[utf8]{inputenc} % allow utf-8 input
\usepackage[T1]{fontenc}    % use 8-bit T1 fonts
\usepackage{hyperref}       % hyperlinks
\usepackage{url}            % simple URL typesetting
\usepackage{booktabs}       % professional-quality tables
\usepackage{amsfonts}       % blackboard math symbols
\usepackage{nicefrac}       % compact symbols for 1/2, etc.
\usepackage{microtype}      % microtypography
\usepackage{lipsum}
\usepackage{color}
\usepackage{algorithm}
\usepackage{algorithmic}
\usepackage{amsmath,amsfonts,amssymb}
\usepackage{subfigure}
\usepackage{lineno}
\usepackage{graphicx}
\usepackage{bm}
\usepackage{subfigure}
\usepackage{setspace}
\usepackage{graphicx}
\usepackage{newtxtext}
\usepackage{newtxmath}
\usepackage{natbib,bm}
\usepackage{siunitx}
\usepackage{amsmath,amssymb,mathtools}
\usepackage{hyperref,lipsum}
\usepackage{booktabs}
\hypersetup{
    colorlinks = true,
    urlcolor   = blue,
    citecolor  = black,
}

\newcommand{\RomanNumeralCaps}[1]
\linenumbers
\graphicspath{{./figs/}}

\usepackage{color}
\definecolor{blue}{rgb}{0, 0.4470, 0.7410}
\definecolor{red}{rgb}{0.8500, 0.1250, 0.0480} 
\definecolor{green}{rgb}{0.4660, 0.6740, 0.1880}

\title{Studying oscillation death in two-dimensional cylinder-airfoil interactions with \\
synchronization-theoretic autoencoder}

\author{
Yuta Ushida$^{[1]}$, Daiki Beppu$^{[1]}$, Yuzuru Kato$^{[2]}$, and Kai Fukami$^{[1,*]}$
\\
\\
1. Department of Aerospace Engineering, Graduate School of Engineering, Tohoku University,\\
Sendai, Miyagi 980-8579, Japan\\
2. Department of Complex and Intelligent Systems, Future University Hakodate\\
  Hokkaido, 041-8655, Japan
~\\
$^{*}$Corresponding author: kfukami1@tohoku.ac.jp
}

\begin{document}
\maketitle

\begin{abstract}

% \vspace{-2mm}
Body-body interactions have long been investigated to leverage the performance of various fluid-based machines.
This study analyzes the flow interaction between a cylinder and a NACA0012 airfoil using direct numerical simulation and machine learning.
We consider a two-dimensional incompressible flow at Reynolds number 100, where the airfoil is positioned in the wake of a cylinder.
By sweeping the parameter space, composed of the distance between the cylinder and the airfoil~$\Delta x$, the relative height~$\Delta y$, and the angle of attack of the airfoil~$\alpha$, four characteristic vortex-shedding regimes are observed.
This study particularly focuses on suppressed vortex shedding at a certain arrangement as oscillation death in coupled oscillators, with the theoretical prediction of flow stability on a machine-learned low-order coordinate.
Independent flows around each body are regarded as a system that exhibits isolated periodic vortex shedding at infinite distance, and interactions appear at smaller distances.
We employ a synchronization-theoretic autoencoder to derive models for two coupled oscillators from an aerodynamic coefficient.
The proposed approach, which enables the assessment of system stability and the prediction of vortical interaction behavior, may pave the way for the analysis of unsteady body-body interactions from the data-driven and synchronization perspectives.

\end{abstract}

\section{Introduction}
\label{sec:intro}

Interactions between bodies in fluid flow are of interest as they ubiquitously emerge in a range of fluid dynamic scenarios.
Fluid mechanicians have admired them throughout theoretical, experimental, and numerical endeavors, while aiming to leverage such interactions to improve the performance of various fluid-based machines~\citep{RN98}.
In this paper, we study two-dimensional cylinder-airfoil interactions to reveal their underlying mechanism with numerical simulations and data-driven reduced-order modeling.

Examples of theoretical analysis of body-body interactions in flow are found in biology, including a dolphin family~\citep{Dolphin2004}, and a bird-formation flight~\citep{Formation_birds, Lobster_drag, May1979, Couzin2005, Nagy2010}.
Lissaman $\&$ Shollenberger~(1970)~\citep{Formation_birds} showed that 25 birds in a V-formation could extend their flight range by 70 percent compared to a lone bird, due to decreased induced drag and its redistribution by the interaction between each bird. 
An object flying by dynamic lift in a fluid gets its lift by creating downward momentum within its span. 
A bird in a certain formation flight can fly in the upwash of neighbors made by each wingtip vortex, so that it can reduce the induced power with the reduction of the lift generated by their wings~\citep{pelican1988}.
Such induced power depends critically on downwash distribution, and it has been revealed that the V-formation is the only one in which the drag can be evenly distributed~\citep{Formation_birds}.

% fish school
Experimental endeavors have also been carried out to study body-body interactions. 
Li {\it et al}.~(2020)~\citep{Li2020} revealed the long-time hypothesis that flying and swimming animals could exploit neighbor-induced flows. 
They developed bio-mimetic fish-like robots that allow them to directly measure the energy consumption associated with swimming together in pairs. 
They constructed two identical robotic fish, 45 cm in length and 800 g in mass. 
Ristroph $\&$ Zhang~(2008)~\citep{PhysRevLett_anomalous} found that it was the leader of a group who enjoyed a significant drag reduction of up to 50 percent, while the downstream flag suffered a drag increase in their experiments on schooling flapping flags. 
They inserted thin, flexible filaments into a flowing soap film to experimentally model the interaction of shape-changing bodies at the high Reynolds number ${Re}\simeq10^4$. 
This counterintuitive inverted drag relationship is rationalized by dissecting the mutual influence of shape and flow in determining drag.

% transition from EFD to CFD
While these experimental studies have provided grounded knowledge from the real world, numerical approaches can yield detailed, flexible simulations and enable rapid exploration. 
The collective motion~\citep{VICSEK201271} of two parallel fish that move freely in both the longitudinal and lateral directions was numerically investigated by 
Lin {\it et al}.~(2026)~\citep{Lin_Liao_Xie_Wu_2026}, focusing on the effects of wavelength and phase difference on their stable arrangement, swimming speed, and energy consumption.
Mizushima $\&$ Suehiro (2005) numerically investigated the instability and transition of flow past two circular cylinders in tandem arrangement.
They performed a direct numerical simulation,
showing that the abrupt change in some physical quantities, such as drag coefficient~$C_D$, lift coefficient~$C_L$, and Strouhal number~$St$, occurs due to the multiple solutions of the basic equations at a Reynolds number $Re=100$.

These numerical and experimental endeavors enable the curation of high-fidelity fluid flow data~\citep{Brunton2020_review, Taira2025_critical}.
% Synchronization theory
To translate such an extensive amount of fluid flow data into practical physics-based insights, one can consider interpreting fluid flow phenomena from the perspective of synchronization theory~\citep{Pikovsky_Rosenblum_Kurths_2001} and dynamical systems~\citep{Marchioro1999}.

For example, the flow over a bluff body and its dynamics can often be described in a low-order, simplified manner. Once its Reynolds number exceeds a certain critical threshold, a steady symmetric wake becomes von Kármán vortex shedding, which is characterized by a Hopf bifurcation. 
This limit cycle movement can be modeled with the Stuart-Landau equation or the van der Pol oscillator~\citep{Karniadakis1992, Dusek_Gal_Fraunie_1994, Sau2009, Noack2003}. 
By describing fluid flows as low-dimensional nonlinear oscillators, it is possible to examine the response of fluid flow to a perturbation, from the transient to the unsteady state, such as turbulence, and its synchronizing characteristics.

However, we barely see such fluid flows with a single body.
In the case that some body exists in a closer region, for example, two cylinders in a side-by-side~\citep{Zhang2022_wake} or staggered arrangement~\citep{Morimoto2021_piv, Alam2016_staggered}, and a cylinder and an airfoil in tandem, a vortex wake shedding can act as a periodic fluid forcing to the downstream body. 
This system consists of the upstream wake oscillator and the downstream body, which may be explained as the coupling of nonlinear oscillators rather than simple linear superposition. 
This coupling system shows extremely diverse dynamics such as phase locking, quasi-periodicity, and amplitude modulation~\citep{Pikovsky_Rosenblum_Kurths_2001}.

An interesting phenomenon in coupling nonlinear oscillators is {\em oscillation death}~\citep[quenching;][]{Pikovsky_Rosenblum_Kurths_2001}. 
It can be observed when the coupling is of a diffusive type with certain conditions on the magnitude of coupling and frequency detuning.
One point becomes stable, and the oscillations in both systems die out due to coupling, which can be predicted via linear stability analysis for low-dimensional problems~\citep{Pikovsky_Rosenblum_Kurths_2001}.
In the context of fluid mechanics, limit-cycle oscillations, such as flutter or von Kármán vortex shedding, can be stopped with the occurrence of oscillation death by introducing body-body interaction~\citep{RAAJ2021103659, Suehiro2005}.
This suggests that oscillation suppression in a high-dimensional fluid flow space could be predicted through synchronization theory if an appropriate low-order coordinate is prepared~\citep{FF2025,brunton2021applying}.
While the choice of such coordinates is often non-trivial, we have found that the incorporation of synchronization theory into nonlinear machine-learning-based coordinate transformation enables tractable prediction of oscillation death for high-dimensional flow phenomena in a reduced-order manner.
This can not only deepen our physical understanding but also offer theoretical pathways to mitigate noise and structural fatigue in advanced aerospace design.

%%%%%%%%%%%%%%%%%%%%%%%%%%%%%%%%%%%%%%%%
\begin{figure}[t]
  \centerline{
  \includegraphics[width=1\textwidth]{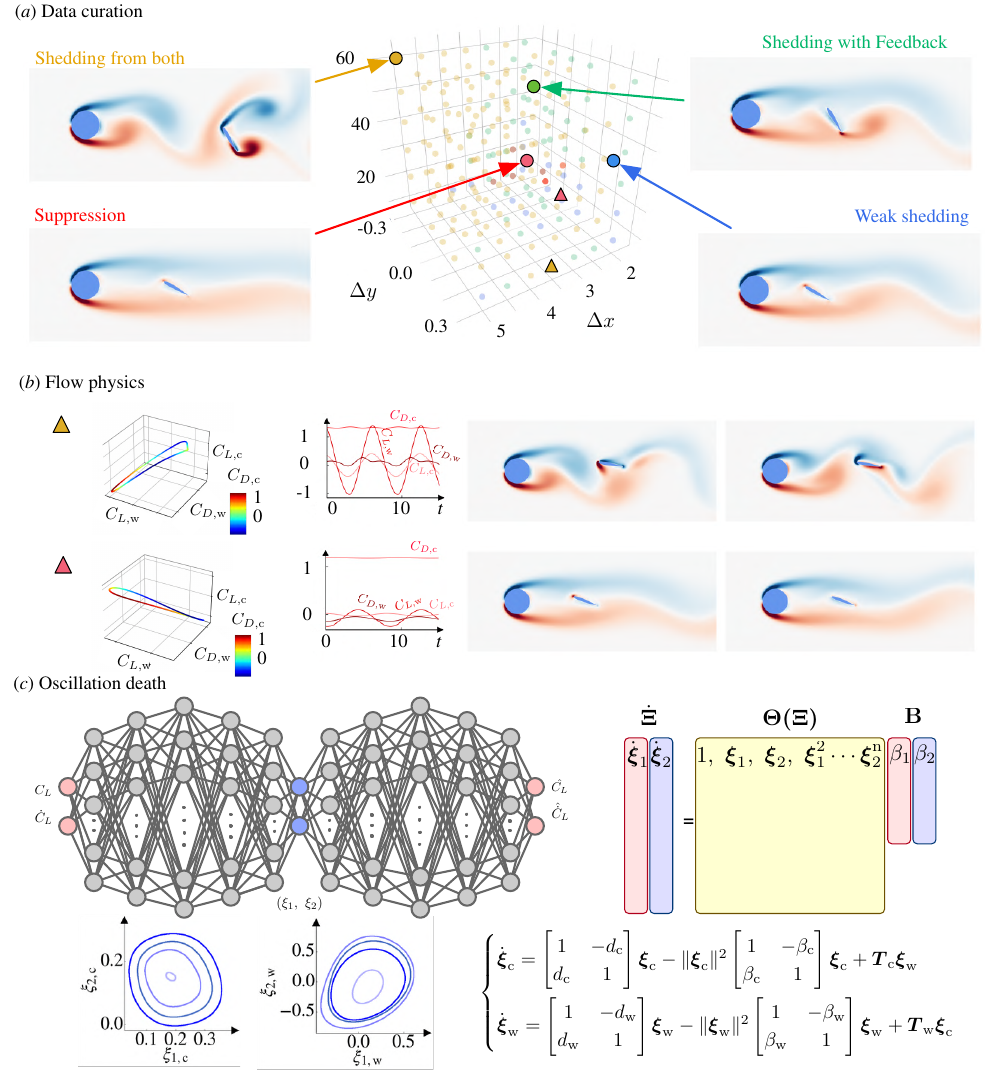}
  }
  \caption{
  Overview of the present study: $(a)$ data curation, $(b)$ study of flow fields and aerodynamic coefficients, and $(c)$ examination of oscillation death.
  }
\label{fig1}
\end{figure}
%%%%%%%%%%%%%%%%%%%%%%%%%%%%%%%%%%%%%%%%

% Overview
In the present work, we consider an interaction between a stationary cylinder and an NACA0012 airfoil at a chord length $c$ or diameter $D$-based Reynolds number of $Re_{\it L} = 100$ ($c$=$D$=$L$), presenting a variety of aerodynamic interaction characteristics. The overview of this study is presented in figure~\ref{fig1}. We first perform a parametric study using direct numerical simulations to examine the influence of the geometric relationship between the cylinder and the airfoil on their interaction. 
Once we categorize the characteristic flow patterns observed from the current dataset, we then examine the effects of each parameter, the distance between the cylinder and the airfoil $\Delta x$, the relative height of the airfoil to the cylinder $\Delta y$, and the angle of attack of the airfoil $\alpha$, on lift and drag for both bodies.
The dynamic mode decomposition~\citep{SCHMID_2010} is then applied to vorticity snapshots, extracting structures that likely contribute to the interaction between the bodies.
To further analyze the coupling of these phenomena, we propose a synchronization-theoretic autoencoder that derives low-order models describing two periodic oscillators from aerodynamic coefficient data. 
We identify the equation for each coupling dynamics and then predict the condition of the oscillation death regime via linear stability analysis.
The remainder of this paper is organized as follows: section 2, method; section 3, results. Concluding remarks are offered in section 4.

% \vspace{-5mm}
\section{Methods}
\label{sec:method}

%%%%%%%%%%%%%%%%%%%%%%%%%%%%%%%%%%%%%%%%
\begin{figure}
  \centerline{
  \includegraphics[width=1\textwidth]{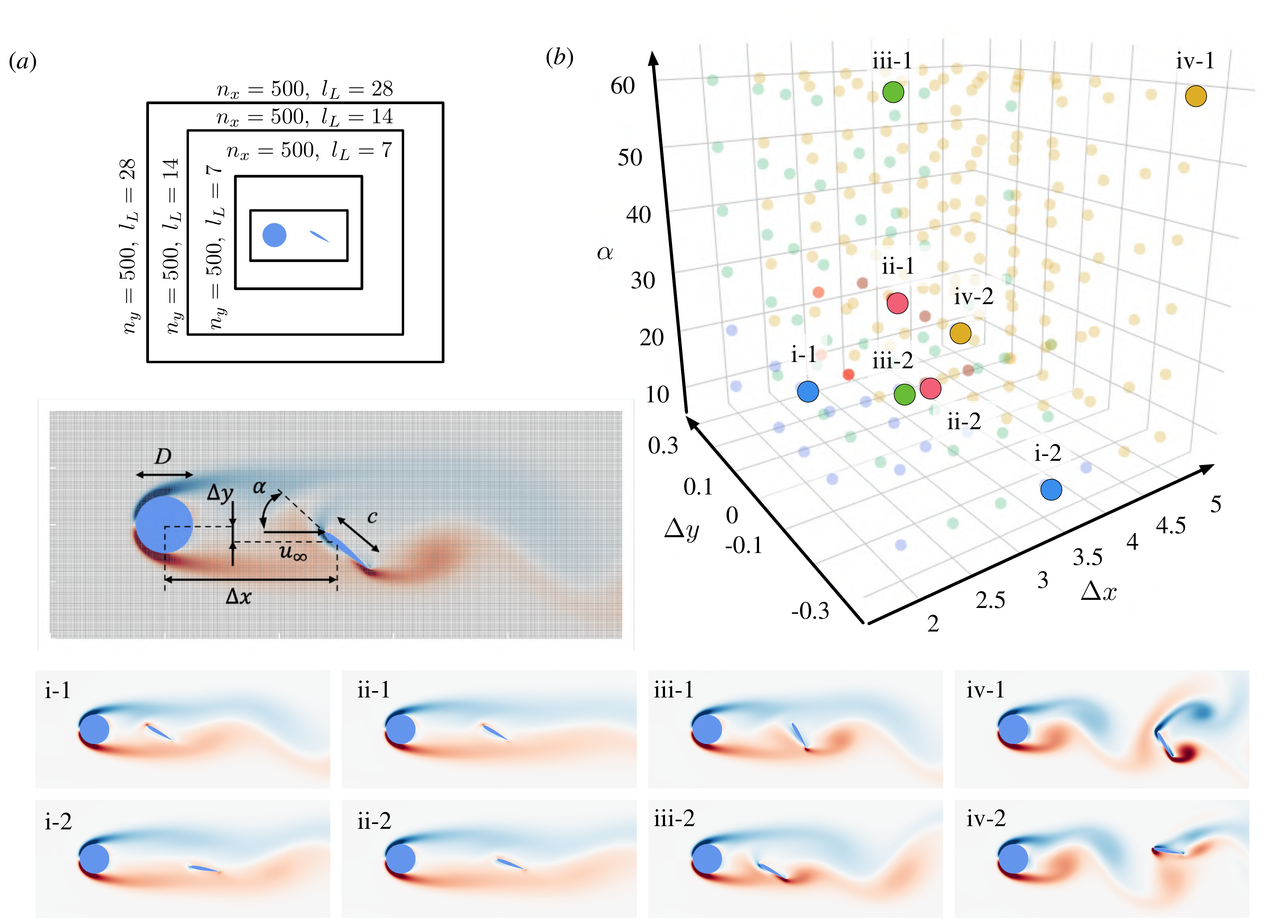}
  }
  \caption{
  $(a)$ The present multi-computational domains and the representative setup with the cylinder and NACA0012 airfoil at $\alpha=40^\circ$ and the distance in the streamwise and vertical directions being $\Delta x = 2$ and $\Delta y = -0.3$, respectively.
  $(b)$ The current space with the eight representative vorticity fields.
  }
\label{fig2}
\end{figure}
%%%%%%%%%%%%%%%%%%%%%%%%%%%%%%%%%%%%%%%%

This study examines cylinder-airfoil interaction flows with numerical simulations and data-driven techniques.
We first study flow characteristics with direct numerical simulations.
We then consider coordinate transformation of the aerodynamic coefficient with phase-inspired autoencoders, and examine oscillation death via linear stability analysis. 
In the following subsection, we present the simulation setup for the cylinder-airfoil interaction flow and its validation/verification study. 
Subsequently, we describe the architecture of the neural network for transforming the coordinates of the aerodynamic coefficients obtained from the simulation. 
We then introduce how the oscillation death is examined with an identification of ordinary differential equations and the linear stability analysis.

\subsection{Simulation setup}
\label{sec:method_DNS}

We perform direct numerical simulations of a cylinder and NACA0012 airfoil at a chord length $c$ or diameter $D$-based Reynolds number of $Re_L = u_\infty L/\nu=100\ (c=D=L)$.
Throughout this paper, we non-dimensionalize all the quantities by the characteristic length $L=D=c$, where $D$ is the cylinder diameter and $c$ is the chord length, and velocities are scaled by the freestream velocity $u_\infty$ and the viscosity $\nu$. 
Diverse configurations for the relative location of the airfoil to the cylinder is considered, as shown in figure~\ref{fig2}.
The vorticity field and force data obtained for each case are analyzed to acquire knowledge on the underlying effect of the body-body interaction between the cylinder and the airfoil. The present flow is parameterized with these variations of $\alpha$, $\Delta x$, and $\Delta y$, as dipicted in figure~\ref{fig2}(a). 
The simulations cover six angles of attack $\alpha = [10,60]^\circ$ at the 35 different locations of the quarter-chord point of the airfoil at $x_0/c\in [2.5,5.0]$ and $y_0/c\in [-0.3,0.3]$. 
The angle of attack of the airfoil $\alpha$ can change the intensity and the frequency of the vortex shedding. 
The distance between the two bodies $\Delta x$ is primally affects the interaction magnitude. 
The relative height of the airfoil to the cylinder $\Delta y$ is also introduced to change where vortex shedding of the cylinder impinges on the airfoil.

%%%%%%%%%%%%%%%%%%%%%%%%%%%%%%%%%%%%%%%%
\begin{figure}
  \centerline{
  \includegraphics[width=0.9\textwidth]{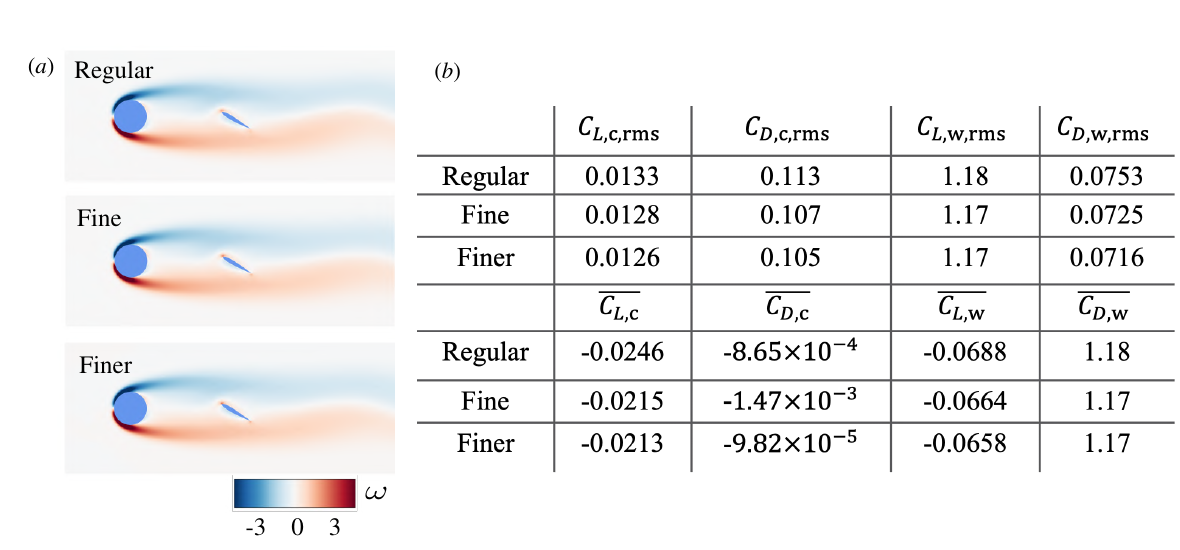}
  }
  \caption{{
  $(a)$ Comparisons of an instantaneous vorticity field $\omega$.
  $(b)$ Comparisons of lift and drag forces (root-mean-squared value and time-averaged value).
  }} % Add CDRMS
\label{fig3}
\end{figure}
%%%%%%%%%%%%%%%%%%%%%%%%%%%%%%%%%%%%%%%%

The immersed boundary projection method is used to numerically simulate the flow~\citep{RN106,RN107}. 
This method is based on a finite volume formulation and incorporates the no-slip boundary condition along the immersed boundary into the projection operation. 
The scheme is second-order accurate in time and has a spatial accuracy of higher than first order in the $L_2$ norm.
We use a multidomain technique to resolve the flow field, while reducing the computational cost.
% The scheme has been validated for several cases~\cite{TAIRA_COLONIUS_2009,Jantzen2014}, and has been found robust and accurate~\cite{RN107}.
Three nested levels of multidomains are used, with the finest level being $(x/c,y/c)\in[-1,6]\times[-1,6]$ and the largest domain being $(x/c,y/c)\in [-11.5,16.5]\times[-11.5,16.5]$ in size, with the center of the cylinder positioned at the origin. 
The drag and lift coefficients are defined as
\begin{equation}
\label{eq:1}
    C_{D,{\rm c}} \equiv \frac{F_{x,{\rm cylinder}}}{(1/2)\rho u_\infty^2 L}\ {\rm ,} \ C_{L,\rm{c}}\equiv\frac{F_{y,{\rm cylinder}}}{(1/2)\rho u_\infty^2 L}\ {\rm ,}\ C_{D,\rm{w}} \equiv \frac{F_{x,{\rm airfoil}}}{(1/2)\rho u_\infty^2 L}\ {\rm and}\ C_{L,\rm{w}}\equiv\frac{F_{y,{\rm airfoil}}}{(1/2)\rho u_\infty^2 L},
\end{equation}
respectively.
Here, the subscripts of $(\cdot)_x$ and $(\cdot)_y$ denote the direction of each body force, $(\cdot)_L$ and $(\cdot)_D$ represent the streamwise and the vertical direction, respectively, and $(\cdot)_c$ and $(\cdot)_w$ show the corresponding body, the cylinder, and the airfoil.
The fluid density is described by $\rho$.
The shedding frequency $f_s$ of each aerodynamic force is nondimensionalized as the Strouhal number; $St\equiv f_s L/u_\infty$, in which $L$ is the characteristic length $L=D=c$.

A grid-convergence study is performed at $\alpha=40^\circ$ with $(x_0/c,y_0/c)=(3,0)$ for grid sizes ranging from $300\times300$ to $700\times700$.
The grid sizes of $300\times300$, $500\times500$, and $700\times700$ are called as {regular}, {fine}, and {finer}, respectively.
The comparison of a representative vorticity snapshot and the aerodynamic coefficients is shown in figure~\ref{fig3}. 
The $500\times500$ grid resolution ensures convergence of the aerodynamic coefficients, capturing the wake structures. 
Thus, the fine grid resolution is deemed adequate to yield accurate results and is used throughout this study.

\subsection{Nonlinear dynamic analysis in a machine-learned latent space}
\label{Chapter2_2_1}

This study characterizes the suppression of vortex shedding due to the body-body interactions using stability analysis and a nonlinear autoencoder.
Let us here briefly explain the rationale for the use of an autoencoder-based coordinate transformation in our study. 
The flow around the cylinder is of interest to the community, exhibiting a tremendous history in fluid mechanics and dynamical systems. 
It has been shown that instabilities occur due to a series of Hopf bifurcations that happen as the flow velocity increases~\citep{Ruelle1971} and create a richer and more complicated fluid nature. 
The Hopf bifurcation was found in the fluid system: the transition from a steady laminar wake to laminar periodic vortex shedding at a Reynolds number of 47~\citep{Zebib1987, Jackson1987}. 
This discovery prompted a long-standing discussion within the community about how a Hopf bifurcation with cubic nonlinearity can be represented in a fluid flow governed by the Navier--Stokes equations with up to quadratic nonlinearity. 
Noack {\it et al.} (2003) \citep{Noack2003} resolved this by separating timescales and using a mean-field model.

We find that the current cylinder-airfoil interaction can be characterized into four representative regimes, as will be discussed later in section~\ref{sec:res}. 
We posit that Hopf bifurcation can occur from one of the regimes to another as the configuration parameters change.
Once such a bifurcation is defined, it is possible to describe it by the form of weakly coupled Stuart--Landau oscillators~\citep{Stuart1958, Landau1944}.

To define the current cylinder-airfoil interaction as such, we consider an autoencoder that transforms the lift coefficient $C_L$ and its time derivative $\dot{C_L}$ of each body, ensuring that the latent dynamics show two coupled Stuart--Landau oscillators. 
Each oscillator corresponds to the cylinder and the airfoil, respectively. 
As will be shown, the phase space of $C_L-\dot{C_L}$ is a natural choice for the input of the autoencoder since they represent the limit-cycle oscillations of the current flows.

\subsubsection{Proposed synchronization-inspired autoencoder}
\label{Chapter2_2_2}

%%%%%%%%%%%%%%%%%%%%%%%%%%%%%%%%%%%%%%%%%%%%%
\begin{figure}
  \centerline{
  \includegraphics[width=0.7\textwidth]{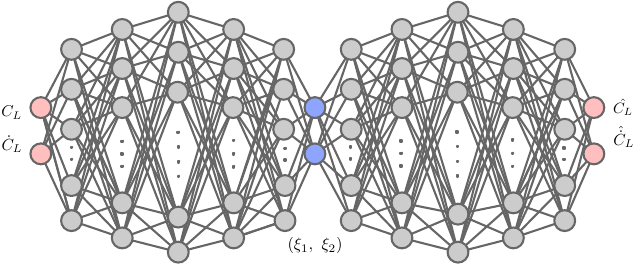}
  }
  % \vspace{-3mm}
  \caption{
    The proposed synchronization-theoretic autoencoder.
  }
\label{fig_mtd1}
\end{figure}
%%%%%%%%%%%%%%%%%%%%%%%%%%%%%%%%%%%%%%%%%%%%%

Hereafter, let us discuss how such an autoencoder is considered by involving synchronization theories.
An autoencoder~\citep{Hinton2006, Fukami2023} is composed of an encoder $f_{\mathrm{enc}} : {\mathbb R}^{d_C} \to {\mathbb R}^{d_{\boldsymbol{\xi}}}$, which transforms 
an input vector ${\boldsymbol{D}}   \in {\mathbb R}^{d_C}$ to
a latent vector ${\boldsymbol{\xi}} \in {\mathbb R}^{d_{\boldsymbol{\xi}}}$,
and a decoder $f_{\mathrm{dec}} : {\mathbb R}^{d_{\boldsymbol{\xi}}} {\to} {\mathbb R}^{d_{X}}$, which approximately reconstructs the original vector $\widehat{\boldsymbol{D}} \in {\mathbb R}^{d_C}$ from the latent vector ${\boldsymbol{\xi}}$. 
Here, $d_{C}$ and $d_{\boldsymbol{\xi}}$ are the dimensions of the input and latent spaces, respectively.
In this study, $d_{\boldsymbol{\xi}}$ and $d_{X}$ are fixed at 2. 

For each coefficient, we use different machine-learning architectures based on a multi-layer perceptron~\citep{Rumelhart1986_backprop}. 
This is because the complexity of the lift-coefficient trajectory for each body varies. 
The lift coefficient of the cylinder initially resembles that of a circle or ellipse, whereas that of the airfoil does not exhibit such a simple trajectory, as will be shown later. 
To employ the coupled Stuart--Landau oscillators for the fitted equations, we need to transform the coordinates of the lift coefficient and its time derivative, which do not show the circle of the noncoupled system, nor the ellipses of the coupled ones. 
Here, we utilize the autoencoder based on the synchronization theory-based loss function. 
Further details on the model configuration are found in the source code~(\url{https://github.com/kfukami/Synch_AE}).

We assume that the oscillator state ${\boldsymbol{D}}(t)$ is sampled at a time interval of $\tau > 0$, which is denoted as ${\boldsymbol{D}}_t$, and transformed to a latent vector ${\boldsymbol{\xi}}_t = f_{\mathrm{enc}}({\boldsymbol{D}}_t)$ at time step $t$ by the autoencoder. 
We design the autoencoder such that the latent variables (components of the two-dimensional latent vector) ${\boldsymbol{\xi}}_t = (\xi_{1,t},\xi_{2,t})$ satisfy 
\begin{gather}
    \boldsymbol{\xi}_{t+\tau} = \boldsymbol{R}(\theta)\boldsymbol{S}\boldsymbol{r}(\omega\tau)\boldsymbol{S}^{-1}\boldsymbol{R}(\theta)^{-1}
        (\boldsymbol{\xi}_{t} - \bar{\boldsymbol{\xi}}) + \bar{\boldsymbol{\xi}},
    \label{eq:2_8}
\end{gather}
where 
\begin{gather}    
    \boldsymbol{R}(\theta)=\begin{pmatrix}
                    \cos{\theta} & -\sin{\theta} \\
                    \sin{\theta} & \cos{\theta}
                    \end{pmatrix},
    \boldsymbol{S}=\begin{pmatrix}
                         a & 0 \\
                         0 & b
                    \end{pmatrix},
    \label{eq:2_9}\\
    \boldsymbol{r}(\omega\tau)=
                    \begin{pmatrix}
                      \cos{\omega\tau} & -\sin{\omega\tau} \\
                      \sin{\omega\tau} & \cos{\omega\tau}
                    \end{pmatrix},
    \label{eq:2_10}\\
    \| \boldsymbol{S}^{-1}\boldsymbol{R}(\theta)^{-1}{\boldsymbol{\xi}_t} \|_2=1,
    \label{eq:2_11}
\end{gather}
$\theta$, $\omega$, $a$, and $b$ are parameters to be learned for the case with interaction, while they, except for $\omega$, are fixed for the case without interaction: independent flows as $\theta=0$, $a=1$, and $b=1$, leaving only the angular frequency $\omega$ as a learnable parameter.
The time-averaged latent vector $\overline{\boldsymbol{\xi}}$ is also used.
How we transform the latent space to a unit circle is explained in equation~(\ref{eq:2_8}).
As their shapes in the latent space can be tilted or non-centered, we may need to perform a coordinate transformation from the latent space to another space.
The operation in equation~(\ref{eq:2_9}) for the coordinate transformation shows that the oscillator state ${\boldsymbol{D}}_t$ is mapped onto an ellipse or a circle in the transformed latent space.
The rotational matrix in equation~(\ref{eq:2_10}) for the latent variable $\boldsymbol{\xi}$ is used such that the transformed latent vector is rotated at a learnable frequency $\omega$ as the oscillator state evolves.
The transformed latent vector is then projected onto a unit circle via equation~(\ref{eq:2_10}).
We design the autoencoder to satisfy equations~(\ref{eq:2_8}-\ref{eq:2_11}) for all $\tau>0$.

The decoder is represented as
\begin{equation}
    \boldsymbol{\widehat{D}} = f_{\mathrm{dec}}({\boldsymbol{\xi}}),
\end{equation}
which takes the normalized latent vector ${\boldsymbol{\xi}}$ as the input and output the oscillator state $\boldsymbol{{\widehat{D}}}$ approximating the original state ${\boldsymbol{D}}$.
Note again that we prepare two autoencoders: one is for a cylinder, and another is for an airfoil.
The aforementioned operations are inspired by~\cite{Yawata2024_chaos, Yawata2025_phase}, who developed a phase autoencoder for data-driven phase-reduction analysis of low-order limit-cycle oscillators.

\subsubsection{Loss functions}
\label{Chapter2_2_3}

To train the autoencoder, we introduce four types of loss functions: reconstruction loss, shape loss, dynamics loss, and center-of-mass loss~\citep{Yawata2024_chaos}, as shown in equation~(\ref{eq:loss_arch}).
Using the above three types of loss functions, we introduce the following total loss function,
\begin{equation}
    L = w_{\mathrm{recon}} L_{\mathrm{recon}} + w_{\mathrm{shape}} L_{\mathrm{shape}} + w_{\mathrm{dyn}} L_{\mathrm{dyn}} + w_{\mathrm{mass}} L_{\mathrm{mass}},
    \label{eq:loss_arch}
\end{equation}
where $w_{\mathrm{recon}}$, $w_{\mathrm{shape}}$, $w_{\mathrm{dyn}}$, and $w_{\mathrm{mass}}$ are weight parameters controlling relative contributions from the individual loss functions.
We minimize this total loss function to train our synchronization-inspired autoencoder.
The reconstruction loss $L_{\rm recon}$ of the autoencoder is the distance between the input data and the reconstructed output,
\begin{equation}
    L_{\rm recon} = {\mathbb E} \left[ \left\|{\boldsymbol{D}}_t - f_{\mathrm{dec}}(f_{\mathrm{enc}}({\boldsymbol{D}}_t)) \right\| \right],
    \label{eq:reconloss}
\end{equation}
where $\| \cdot \|$ represents the Frobenius norm and ${\mathbb E}[ g({\boldsymbol{Q}}_t) ] = \sqrt{(1/T_{\rm max}) \sum_{t=0}^{T_{\rm max}-1} g({\boldsymbol{Q}}_t)^2}$ represents the root-mean-squared value of a quantity $g({\boldsymbol{Q}}_t)$ over the input data $\{ {\boldsymbol{Q}}_t | t=1, ..., T_{\rm max} \}$ of length $T_{\rm max}$.
Here, $T_{\rm max}$ is the maximum number of inputs that can be obtained from a single orbit, determined by the simulation length and its computation step width.

The shape loss $L_{\rm shape}$ is added to guide the latent space to be a circle or an ellipse, depending on whether or not interaction occurs,
\begin{equation}
    L_{\mathrm{shape}} = \mathbb{E} \left[ \left\| \sum^{2}_{i=1}(\bm{R}^{-1} \bm{S}^{-1} \bm{D}_{t})_i^{2} - 1 \right\| \right],
    \label{eq:shapeloss}
\end{equation}
where the latent variables are summed componentwise over the $i$th component. The same matrices of $\boldsymbol{R}$ and $\boldsymbol{S}$ defined in equations~(\ref{eq:2_8}) and (\ref{eq:2_9}) are used, enforcing the transformed latent trajectory to be a circle. 
The independent dynamics: the flow past a cylinder and the flow past an airfoil are exhibited as circles, and the deformation is observed with interaction.

The dynamics loss $L_{\rm dyn}$ controls the properties of the latent space.
We introduce the following loss function characterizing the temporal prediction errors for the latent variables,
\begin{gather}
     L_{\rm dyn} = 
    {\mathbb E}\left[\|f_{\mathrm{enc}}({\boldsymbol{D}}_{t+\Delta t})-\boldsymbol{r}(\omega\Delta t)\boldsymbol{S}^{-1}\boldsymbol{R}^{-1}(f_{\mathrm{enc}}({\boldsymbol{D}}_t)) \|
    \right],  
\end{gather}
where $\Delta t$ is a user-defined time step.
This $\Delta t$ needs to be appropriately scaled to take account of the period $T$ of the limit cycle.
In training, $\Delta t$ is fixed, but it is expected that equations~(\ref{eq:2_9})-(\ref{eq:2_11}) are satisfied for any $\tau>0$ if trained with sufficient data. 
This study uses $\Delta t =0.2$, achieving a stable learning process.

The loss function $L_{\rm shape}$ possesses a local solution that represents a state where the equation identification does not work: the center of the ellipse for the interaction is not at the origin in the latent space. 
If we have the offset of the latent trajectory, the interaction term in the equations~\ref{eq:integrated_system} that describe the interaction has a constant term to make the identification accurate, described in detail in the next subsection~\ref{Chapter2_3}. The constant term corresponds to the nonphysical exponential increase of the latent variable.
In response, we further introduce the following loss function to avoid such an inappropriate solution,
\begin{align}
    L_{\mathrm{mass}}
     &=
    {\mathbb E} \left[
      \sum^2_{i=1}\boldsymbol{\xi}_i
	\right],
\end{align}
which characterizes the center of mass of the latent variables.
The weighting parameters are set to be $(w_{\rm recon},w_{\rm shape},w_{\rm dyn},w_{\rm mass})=(1.00, 0.100, 0.100, 1.00\times 10^{-2})$ and $(1.00,10.0,1.00\times 10^{-2}, 1.00\times 10^{-2})$ for the cylinder and the airfoil, respectively, based on our preliminary studies.

\subsection{Examining oscillation death with synchronization theory}
\label{Chapter2_3}
Once a circle for the non-interaction cases and an ellipse for the interaction cases are obtained, we identify the model equation to represent each shape based on the Stuart--Landau oscillator. 
The model equation is described as,
\begin{equation}
\left\{
\begin{aligned}
\dot{\xi}_{1,{\rm c}} &= \xi_{1,{\rm c}} - d_{\mathrm{c}} \xi_{2,{\rm c}} - (\xi_{1,{\rm c}}^2 + \xi_{2,{\rm c}}^2)(\xi_{1,{\rm c}}-\beta_c \xi_{2,{\rm c}}) + (T_{{\rm c},11} \xi_{1,{\rm w}}+T_{{\rm c},12} \xi_{2,{\rm w}}) \\
\dot{\xi}_{2,{\rm c}} &= \xi_{2,{\rm c}}+ d_{\mathrm{c}} \xi_{1,{\rm c}} - (\xi_{1,{\rm c}}^2 + \xi_{2,{\rm c}}^2)(\xi_{2,{\rm c}} + \beta_c \xi_{1,{\rm c}}) + (T_{{\rm c},21} \xi_{1,{\rm w}}+T_{{\rm c},22} \xi_{2,{\rm w}}) \\
\dot{\xi}_{1,{\rm w}} &= \xi_{1,{\rm w}} - d_{\rm w} \xi_{2,{\rm w}} - (\xi_{1,{\rm w}}^2 + \xi_{2,{\rm w}}^2)(\xi_{1,{\rm w}}-\beta_{\rm w} \xi_{2,{\rm w}}) + (T_{{\rm w},11} \xi_{1,{\rm c}}+T_{{\rm w},12} \xi_{2,{\rm c}}) \\
\dot{\xi}_{2,{\rm w}} &= \xi_{2,{\rm w}} + d_{\rm w} \xi_{1,{\rm w}} - (\xi_{1,{\rm w}}^2 + \xi_{2,{\rm w}}^2)(\xi_{2,{\rm w}} + \beta_{\rm w} \xi_{1,{\rm w}}) + (T_{{\rm w},21} \xi_{1,{\rm c}}+T_{{\rm w},22} \xi_{2,{\rm c}})
\end{aligned}
\right.
\label{eq:integrated_system}
\end{equation}
where subscript ${(\cdot)}_\mathrm{c}$ and ${(\cdot)}_{\mathrm{w}}$ denote the cylinder and the airfoil trajectory, respectively.
The interaction magnitude and direction are expressed in $T$, showing how the latent vector affects the other latent vector, and the subscript shows the row and column numbers of each matrix component. 
The above equations are also expressed in matrix form,
\begin{equation}
\left\{
\begin{aligned}
  \dot{\boldsymbol{\xi}}_{\rm c} &= \boldsymbol{f}_{\rm c} (\boldsymbol{\xi}_{\rm c})
  + \boldsymbol{T}_{\rm c}\boldsymbol{\xi}_{\rm w} \label{eq:int_all}\\
  \dot{\boldsymbol{\xi}}_{\rm w} &= \boldsymbol{f}_{\rm w} (\boldsymbol{\xi}_{\rm w})
  + \boldsymbol{T}_{\rm w}\boldsymbol{\xi}_{\rm c}
\end{aligned}
\right.
\end{equation}
where $\boldsymbol{\xi}_{\mathrm{c}}$, $\boldsymbol{\xi}_{\mathrm{w}}$, $\boldsymbol{T}_{\mathrm{c}}$, $\boldsymbol{T}_{\mathrm{w}}$ are written as,
\begin{gather}
    \boldsymbol{\xi}_{\mathrm{c}}=\begin{pmatrix}
                    \xi_{1, {\mathrm{c}}} \\
                    \xi_{2, {\mathrm{c}}} 
                    \end{pmatrix},
    \boldsymbol{\xi}_{\mathrm{w}}=\begin{pmatrix}
                    \xi_{1, {\mathrm{w}}} \\
                    \xi_{2, {\mathrm{w}}} 
                    \end{pmatrix},
    \boldsymbol{T}_{\mathrm{c}}=
                    \begin{pmatrix}
                      T_{\mathrm{c}, 11} & T_{\mathrm{c}, 12} \\
                      T_{\mathrm{c}, 21} & T_{\mathrm{c}, 22}
                    \end{pmatrix},
    \boldsymbol{T}_{\mathrm{w}}=
                    \begin{pmatrix}
                      T_{\mathrm{w}, 11} & T_{\mathrm{w}, 12} \\
                      T_{\mathrm{w}, 21} & T_{\mathrm{w}, 22}
                    \end{pmatrix},
    \label{eq:2_21d}
\end{gather}
 respectively.
We identify the equation with linear regressions; that is, $d_{\mathrm{c}}$, $d_{\mathrm{w}}$, $\boldsymbol{T}_{\mathrm{c}}$, and $\boldsymbol{T}_{\mathrm{w}}$ are determined.

To avoid ill-conditioning and over-parameterization during the regression process, the identification of the coefficients for without-interaction and with-interaction cases is performed sequentially rather than simultaneously.
This leverages the principles of sparse identification of nonlinear dynamics~\cite[SINDy;][]{Brunton2016_SINDy}.
First, data corresponding to the isolated cases without interaction are prepared via a synchronization-theoretic autoencoder. 
At this stage, the phase portrait forms a perfect unit circle. 
Let $\dot{\boldsymbol{\Xi}}$ be the time derivatives of the latent vectors, $\boldsymbol{\Theta}$ be the library of candidate nonlinear functions derived from the Stuart--Landau terms, and $\boldsymbol{C}$ be the coefficient matrix. 
By subtracting any analytically known terms $\boldsymbol{\Theta}_k \boldsymbol{C}_k$ from the total derivative $\dot{\boldsymbol{\Xi}}$, the coefficients for the isolated single body $\boldsymbol{C}_s$, which depend on each dynamic behavior of interest, are identified via linear regression such that
\begin{equation}
    \dot{\boldsymbol{\Xi}} - \boldsymbol{\Theta}_k \boldsymbol{C}_k = \boldsymbol{\Theta}_s \boldsymbol{C}_s.
    \label{eq:isolatedSINDy}
\end{equation}
{
When we perform the system identification for the first equation in equation~\ref{eq:integrated_system}, we subtract $\xi_{1,\rm{c}}$ and $(\xi^2_{1,\rm{c}} + \xi^2_{2,\rm{c}})\xi^2_{1,\rm{c}}$ from the time derivative of the data~$\dot{\boldsymbol{\Xi}}$. 
The coefficients~$d_c$ and $\beta_c$ are then identified by comparing the subtracted time derivative of the data~$\dot{\boldsymbol{\Xi}} - \boldsymbol{\Theta}_k \boldsymbol{C}_k$ and the remaining data~$\boldsymbol{\Theta}_s \boldsymbol{C}_s$.
}

Once the base dynamics for the single cylinder or wing $\boldsymbol{C}_s$ are successfully identified, the interaction cases are evaluated. 
The methodology mathematically subtracts out both the known base terms and the newly identified single-body terms from the observed time derivative. 
What remains in the residual is hence purely driven by the interaction dynamics. 
The interactive coefficients $\boldsymbol{C}_i$ which contain the target parameters $\boldsymbol{T}_{\mathrm{c}}$ and $\boldsymbol{T}_{\mathrm{w}}$, are then identified by regressing this residual against the interaction library $\boldsymbol{\Theta}_i$,
\begin{equation}
    \dot{\boldsymbol{\Xi}} - \boldsymbol{\Theta}_k \boldsymbol{C}_k - \boldsymbol{\Theta}_s \boldsymbol{C}_s = \boldsymbol{\Theta}_i \boldsymbol{C}_i.
\end{equation}

The linear regression steps described above are performed by solving a least-squares optimization problem. 
The optimal coefficient matrix $\boldsymbol{C}^*$ is found by minimizing the $L_2$ norm of the error between the measured temporal derivatives and the derivatives predicted by the constructed function library,
\begin{equation}
    \boldsymbol{C}^* = \text{argmin}_{\boldsymbol{C}} ||\dot{\boldsymbol{\Xi}} - \boldsymbol{\Theta}\boldsymbol{C}||_2.
\end{equation}

This sequential approach ensures that the fundamental topological features, the limit cycles, are preserved while accurately capturing the complex, spatially dependent cross-coupling between the bodies in the fluid flows, this time only from the lift coefficient. Note that this scheme is applicable for any variables that represent the dynamics.

% \vspace{-6mm}
\section{Results}
\label{sec:res}

\subsection{Vortical pattern and aerodynamic responses}

%%%%%%%%%%%%%%%%%%%%%%%%%%%%%%%%%%%%%%%%
\begin{figure}
  \centerline{
  \includegraphics[width=1\textwidth]{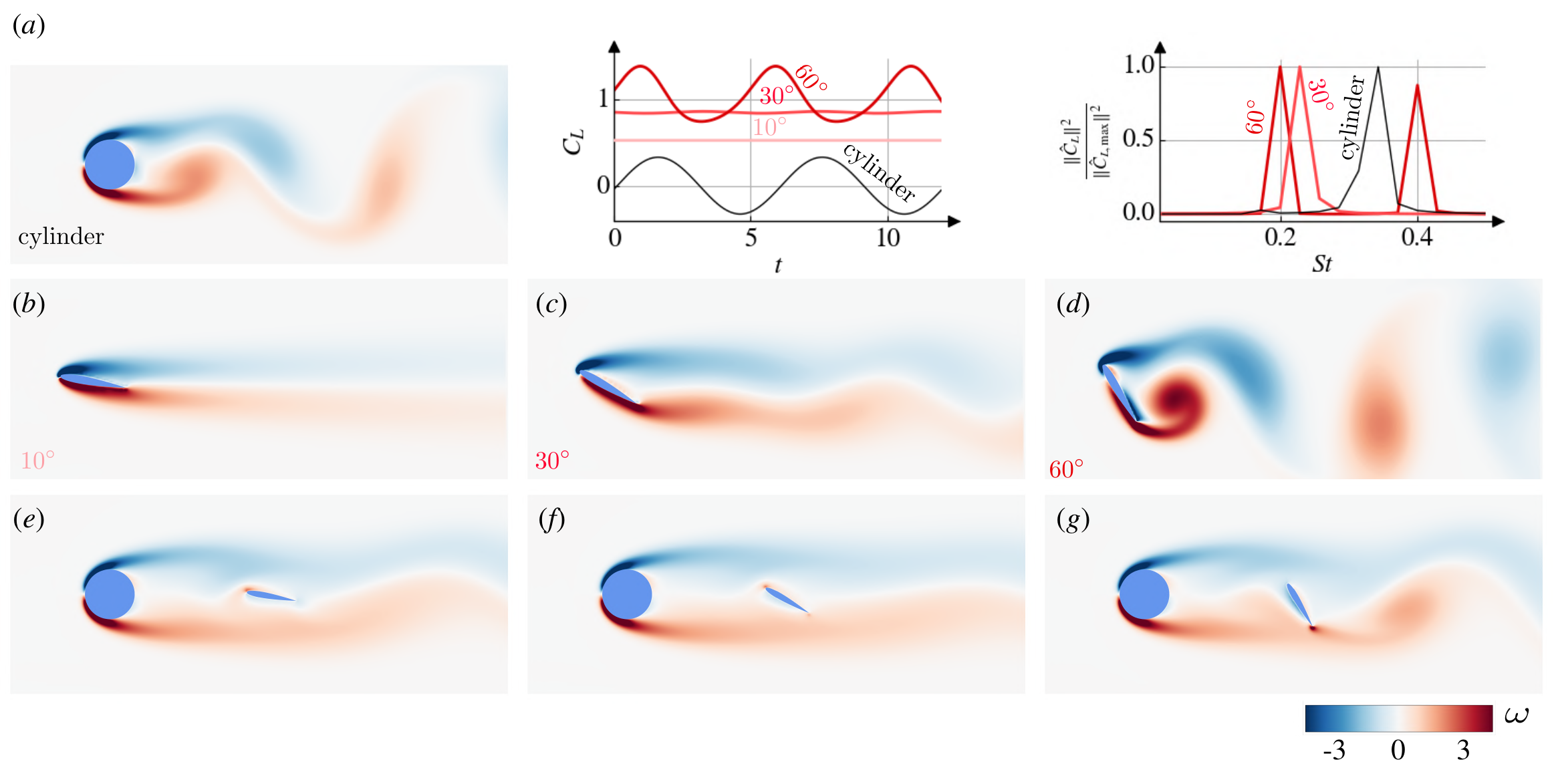}
  }
  \caption{
  Comparison between the single-body flows and the cylinder-airfoil interaction flows.
  Representative vorticity fields with $(a)$ a cylinder and an airfoil at $\alpha =$ $(b)$ $10^\circ$, $(c)$ $30^\circ$, and $(d)$ $60^\circ$.
  The time trace of the lift coefficient $C_L$ and its spectrum are also presented in the top right.
  The cylinder-airfoil interaction flows with $\alpha =$ $(e)$ $10^\circ$, $(f)$ $30^\circ$, and $(g)$ $60^\circ$ are also presented.  
  }
\label{fig4}
\end{figure}
%%%%%%%%%%%%%%%%%%%%%%%%%%%%%%%%%%%%%%%%

Let us examine the flow fields and aerodynamic coefficients obtained from the present simulations.
We begin with the comparison of cases with and without interaction, as summarized in figure~\ref{fig4}. 
For the cylinder flow, von Kármán vortex shedding from a cylinder is observed at the current Reynolds number of 100, as presented in figure~\ref{fig4}$(a)$.
The time series of the cylinder lift coefficient in figure~\ref{fig4}$(b)$ shows a symmetric history, and its time-averaged value is zero due to its symmetric shape and no interacting body upstream or in its wake.
The lift fluctuation is approximately 0.3, which is in agreement with previous studies~\citep{Monkewitz1996,hasegawa2020cnn}. 
There exists one dominant frequency at $St \simeq 0.3$, corresponding to the frequency of the von Kármán vortex shedding from the cylinder.

Let us then consider a flow around an airfoil.
The flow at $\alpha=10^\circ$ is steady, as evidenced by the lift trace in figure~\ref{fig4}$(b)$ and the vorticity field in figure~\ref{fig4}$(d)$.
The von Kármán vortex shedding from an airfoil is then observed for $\alpha \ge30^\circ$~\citep{Fukami2024_attenuation, Choi2015_airfoil}.
For the wake at $\alpha=30^\circ$, the vortex shedding does not appear in the closer region from the airfoil, but evolves as moving toward downstream, as depicted in figure~\ref{fig4}$(e)$. 
At $\alpha = 60^\circ$, the shedding with a larger magnitude of vorticity appears in the wake, as visualized in figure~\ref{fig4}$(f)$.
Correspondingly, the lift fluctuation becomes larger than that at lower $\alpha$.

The frequency spectra of the lift coefficient in figure~\ref{fig4}$(c)$ show that there are two peaks for $\alpha=60^\circ$.
These two peaks are also observed at $\alpha=50^\circ$, although not shown.
This is due to the early evolution of the von Kármán vortex street. 
The negative vorticity region stands out on the suction side of the airfoil at one moment in a period.
This negative vorticity corresponds to the timing of minimum lift, reducing the lift in general~\citep{Godavarthi2023_optimal, Fukami2024_attenuation}. 
%揚力の減少、速度の減少

%%%%%%%%%%%%%%%%%%%%%%%%%%%%%%%%%%%%%%%%
\begin{figure}
  \centerline{
  \includegraphics[width=1\textwidth]{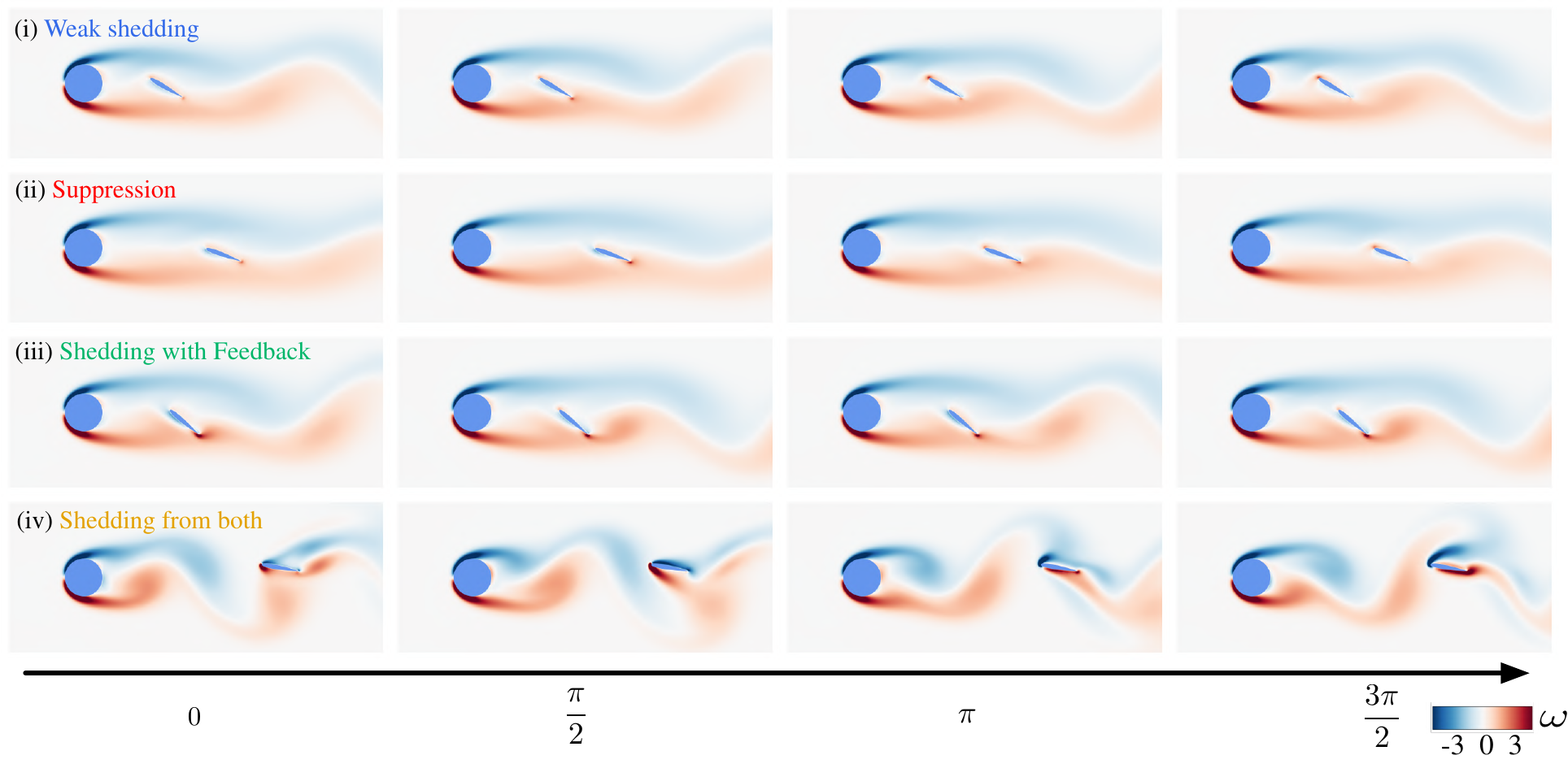}
  }
  \caption{
  Representative vorticity fields for the present four characteristic regimes.
  }
\label{fig_fourch}
\end{figure}
%%%%%%%%%%%%%%%%%%%%%%%%%%%%%%%%%%%%%%%%

% \subsection{Flow-pattern classification}

%%% Add a figure for expressing below: already discussed

The current flow fields can be categorized into four regimes, depending on the characteristics of vortex shedding.
We refer to these four types of observed flow regimes as: 1) {\it weak shedding}, 2) {\it suppression}, 3) {\it shedding with feedback}, and 4) {\it shedding from both}, as presented in figure~\ref{fig_fourch}.
These wake modes are classified mainly based on the flow structure of the wake around the wing and between the cylinder and the airfoil, which can further be evident from vortex-shedding frequency, aerodynamic responses, and dynamic modal structures. 
Although their regime-dependent characteristics will be discussed in detail based on their dependence on the geometric parameters $\alpha$, $\Delta x$, and $\Delta y$ later, let us here briefly introduce their generic flow behaviors.

% Details of four classifications
When $\alpha$ and $\Delta x$ are small, the flow fields present a wake with a single dominant frequency and almost act as a single bluff-body wake, as the cylinder and the airfoil are quite close to each other, as depicted in figures~\ref{fig2}$\rm (i)$ and \ref{fig_fourch}$\rm (i)$.
We refer to this flow regime, presenting shedding from the airfoil and a small fluctuation between the bodies, as {\em weak shedding}.
Both the lift and drag of the cylinder and the airfoil are smaller than those presenting a stronger wake shedding in other categorized regimes. 
This flow regime is sensitive to the choice of the vertical distance $\Delta y$, which determines how the cylinder wake impinges on the airfoil, which will be discussed later in section~\ref{sec:dy}.

The airfoil wake does not possess its own shedding when $\alpha$ is small, as discussed above.
While the cases with small $\Delta x$ show some weak fluctuations in the wake, there exists an optimal spot with respect to $\Delta x$ and $\alpha$ in which the shedding from the bodies is greatly suppressed due to their strong interactions.
This flow regime is called {\em suppression}, whose representative vorticity fields are presented in figures~\ref{fig2}$\rm  (ii)$ and \ref{fig_fourch}$\rm (ii)$.
We particularly focus on this flow pattern and examine the causes of this suppression phenomenon through the lens of oscillation death in the latter section.

By gradually increasing $\alpha$ or decreasing $\Delta x$, unique interactive characteristics between the cylinder and the airfoil emerge.
That is, the flow around the pressure side of the airfoil communicates with the structure with positive vorticity shed by the cylinder, thereby exhibiting feedback from the wing to the cylinder.
We group them as {\em shedding with feedback}.
Example vorticity snapshots are shown in figures~\ref{fig4}$\rm (iii)$ and \ref{fig_fourch}$\rm (iii)$.
Corresponding to stronger shedding in a flow, the magnitude of aerodynamic coefficients is generally larger than in the cases of {\em weak shedding} and {\em suppression}.

As the parameters $\alpha$, $\Delta x$, or $|\Delta y|$ increase and are larger than those of other classifications, the interaction between the bodies becomes weaker in general, presenting von Kármán vortex shedding from both the cylinder and the airfoil as depicted in figures~\ref{fig4}$\rm (iv)$ and \ref{fig_fourch}$\rm (iv)$.
In this classification of {\em shedding from both}, two bodies behave as if they were bodies in isolated flows, although the airfoil is affected by the wake shedding from the cylinder. 
This is similar to aerodynamic scenarios in which a vortex gust is introduced upstream of an airfoil~\citep{jones2022physics,Kaifukami_Re5000,leung2026data}. 
Their wake behaviors vary greatly with different $\alpha$, $\Delta x$, or especially $\Delta y$, which will be discussed in section~\ref{sec:dy}.

% \vspace{-2mm}
\subsubsection{The influence of angle of attack \texorpdfstring{$\alpha$}{alpha}}
\label{sec:alpha}

%%%%%%%%%%%%%%%%%%%%%%%%%%%%%%%%%%%%%%%%
\begin{figure}[t]
  \centerline{
  \includegraphics[width=0.95\textwidth]{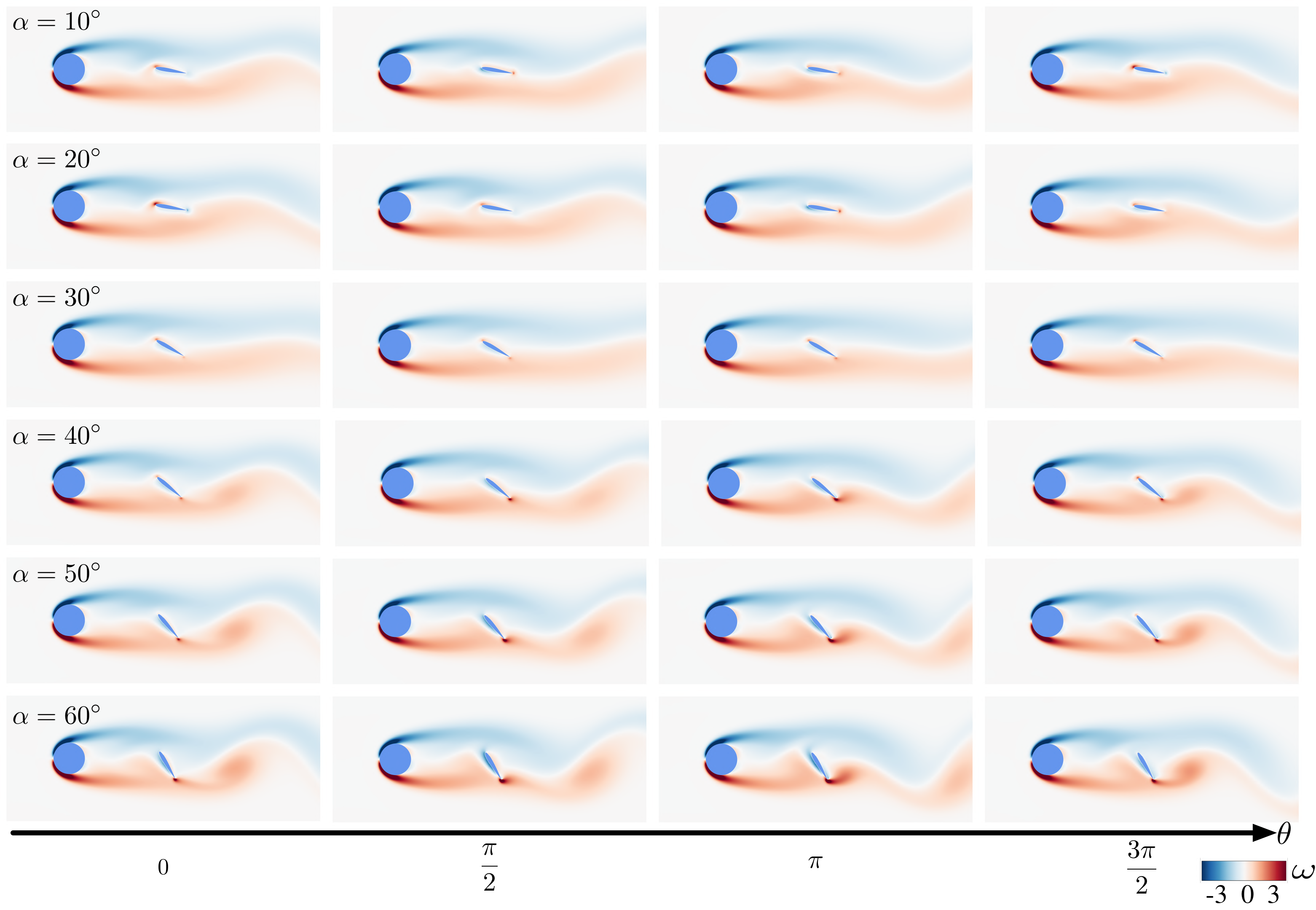}
  }
  \caption{
  The dependence of the vorticity field $\omega$ on the angle of attack of the airfoil $\alpha$ over the phase $\theta$.
  The phase is defined for each periodic time series of the vorticity field.
  % $\theta \equiv 0$ such that $C'_L (\coloneq C_L -\overline{C}_L )=0$.
  The cases for $(\Delta x,\Delta y)=(3,0)$ with $\alpha = [10,60]^\circ$ are shown.
  }
\label{fig5}
\end{figure}
%%%%%%%%%%%%%%%%%%%%%%%%%%%%%%%%%%%%%%%%

%%%%%%%%%%%%%%%%%%%%%%%%%%%%%%%%%%%%%%%%
\begin{figure}
  \centerline{
  \includegraphics[width=0.9\textwidth]{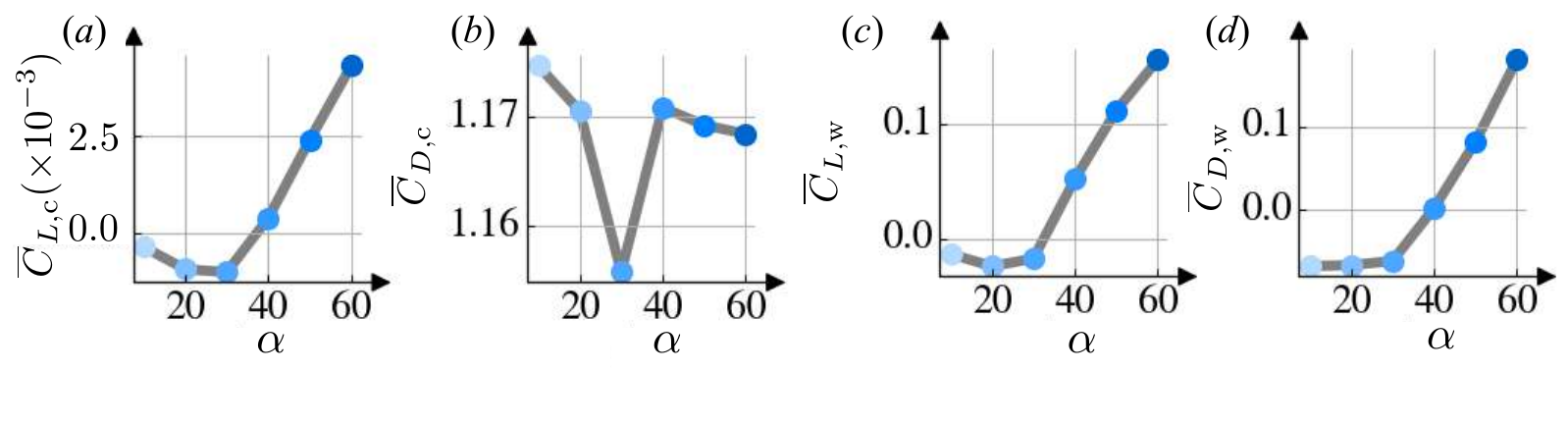}
  }
  \caption{
  The dependence of aerodynamic coefficients $(a)$ $C_{L,\rm{c}}$, $(b)$ $C_{D,\rm{c}}$, $(c)$ $C_{L,\rm{w}}$, and $(d)$ $C_{D,\rm{w}}$ on the angle of attack $\alpha$. 
  The subscripts c and w denote the cylinder and the airfoil, respectively.
  The cases for $(\Delta x,\Delta y)=(3,0)$ with $\alpha = [10,60]^\circ$ are shown.
  }
\label{fig6}
\end{figure}
%%%%%%%%%%%%%%%%%%%%%%%%%%%%%%%%%%%%%%%%

% \vspace{-2mm}
Here, we examine the influence of the angle of attack~$\alpha$ based on the flow fields and lift coefficient.
As representative examples, the cases with $(\Delta x,\Delta y)=(3,0)$ for $\alpha=[10,60]^\circ$ are considered in figure~\ref{fig5}.
Since all the flows studied herein present periodic dynamics, we compare the flow fields over time based on the phase $\theta(t) = \tan^{-1}(C_L^\prime (t)/C_L(t))$. 
At $\alpha = 10^\circ$ and $20^\circ$, the aforeintroduced {\em weak shedding} is observed due to the interaction between the cylinder shedding and the airfoil.

Over the period, the positive vorticity is alternately observed around the leading and trailing edges.
The one around the leading edge is caused by the interaction with the negative vorticity from the cylinder shedding, which uniquely emerges in the interactive flow and not in the isolated flow.
These unique signatures are also highlighted in the time-averaged aerodynamic coefficients, as presented in figure~\ref{fig6}.
For $\alpha \leq 30^\circ$, the lift and drag coefficients of the airfoil report negative values.

%%%%%%%%%%%%%%%%%%%%%%%%%%%%%%%%%%%%%%%%
\begin{figure}[t]
  \centerline{
  \includegraphics[width=0.9\textwidth]{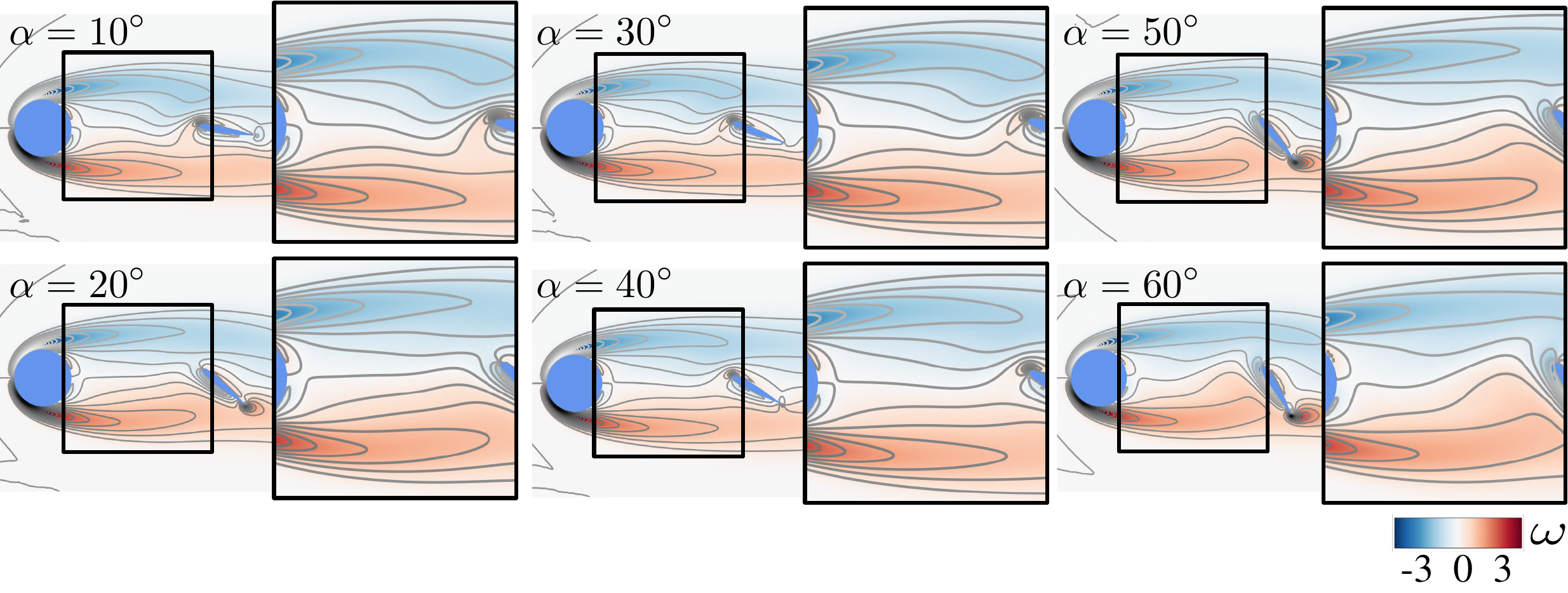}
  }
  \caption{
  Zoomed-in view of the flow between a cylinder and an airfoil across different angles of attack for the cases with $(\Delta x,\Delta y)=(3,0)$ at $\theta={\pi}/{2}$. 
  }
\label{fig:alpha_fluctuation_alpha}
\end{figure}
%%%%%%%%%%%%%%%%%%%%%%%%%%%%%%%%%%%%%%%%

At $\alpha=30^\circ$, the shedding from the airfoil is further suppressed than at the lower angles of attack with respect to vortex size and the magnitude of vorticity, presenting the pattern of {\em suppression}. 
Due to the strong interaction, the fluctuation between the cylinder and the airfoil becomes weaker compared to other angles of attack, as shown in figure~\ref{fig:alpha_fluctuation_alpha}.
This contributes to the extensive decrease in the time-averaged drag coefficient of the cylinder $\overline{C}_{D,\rm c}$, as presented in figure~\ref{fig6}$(b)$.
This angle of attack is hence the optimal angle in the cases considered herein to suppress the shedding.

The airfoil shedding becomes stronger at higher angles of attack for $\alpha \geq40^\circ$.
The feedback vorticity on the pressure side of the airfoil is observed in this range of $\alpha$, categorized as {\em shedding with feedback}.
Since the influence of the angle of attack is more dominant than that of {\em weak shedding} and {\em suppression}, all of the aerodynamic coefficients except for the drag coefficient of the cylinder become larger by increasing the angle of attack, as shown in figure~\ref{fig6}.

% \vspace{-3mm}
\subsubsection{The influence of streamwise distance \texorpdfstring{$\Delta x$}{Delta x}}
\label{sec:dx}

%%%%%%%%%%%%%%%%%%%%%%%%%%%%%%%%%%%%%%%%
\begin{figure}[t]
  \centerline{
  \includegraphics[width=0.9\textwidth]{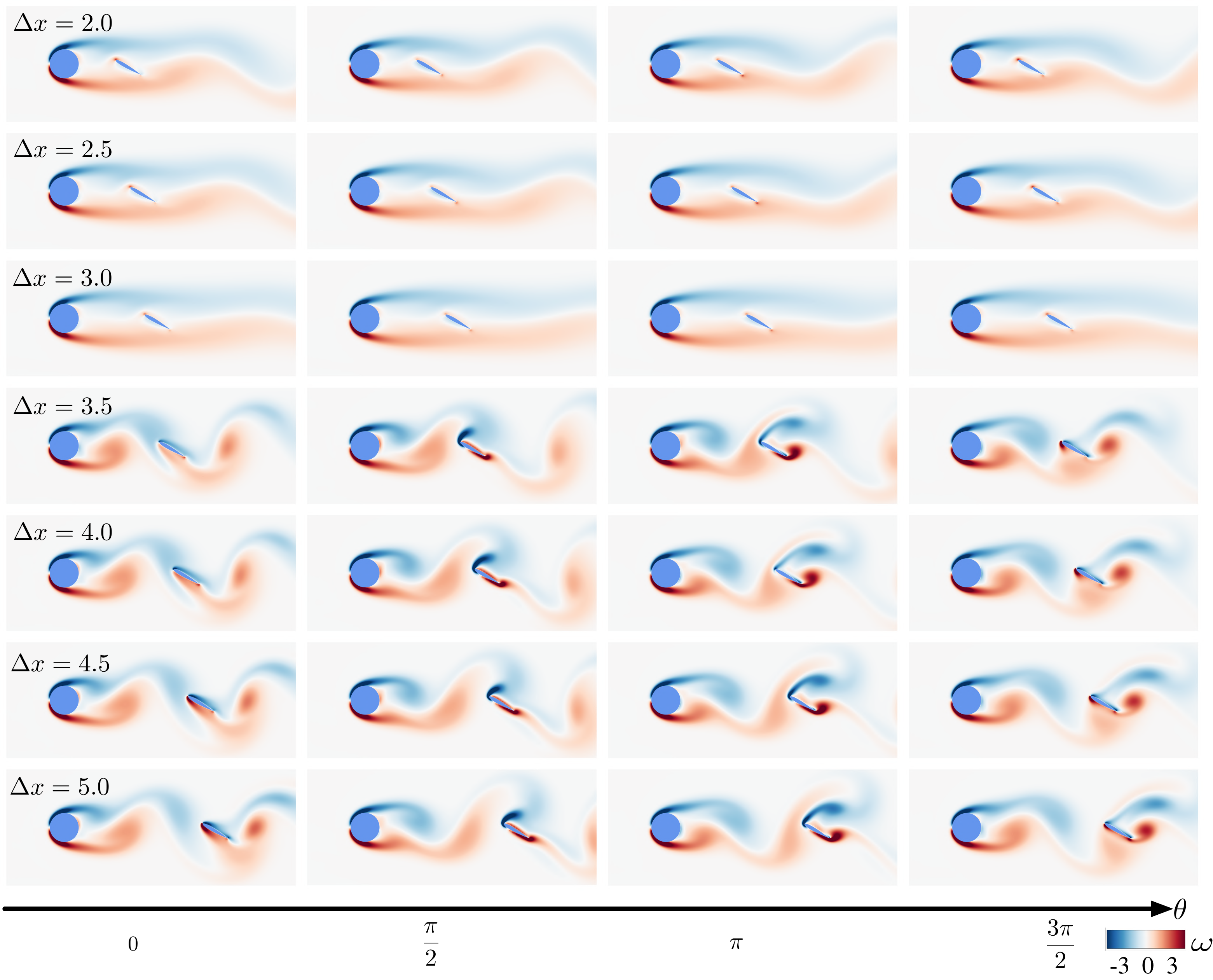}
  }
  \caption{
  The dependence of the vorticity field $\omega$ on the streamwise distance $\Delta x$ over the phase $\theta$.
  The phase is defined for each periodic time series of the vorticity field.
  % $\theta \equiv 0$ such that $C'_L (\coloneq C_L -\overline{C}_L )=0$.
  The cases for $(\alpha,\Delta y)=(30^\circ,0)$ with $\Delta x = [2.0,5.0]$ are shown.  
  }
  % \vspace{-10mm}
\label{fig9}
\end{figure}
%%%%%%%%%%%%%%%%%%%%%%%%%%%%%%%%%%%%%%%%

%%%%%%%%%%%%%%%%%%%%%%%%%%%%%%%%%%%%%%%%
\begin{figure}
  \centerline{
  \includegraphics[width=0.9\textwidth]{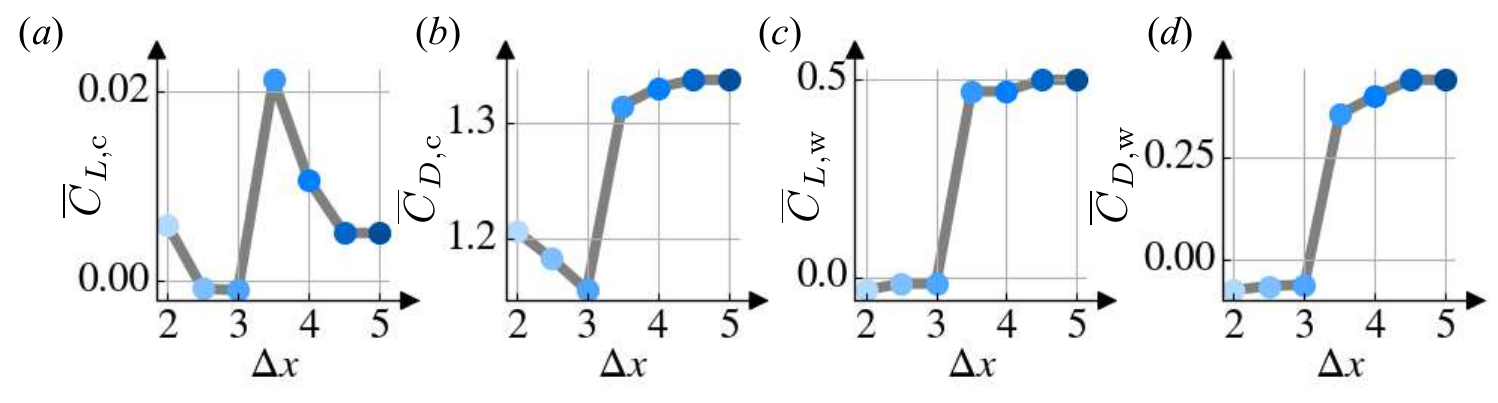}
  }
  \caption{
  The dependence of aerodynamic coefficients $(a)$ $C_{L,\rm{c}}$, $(b)$ $C_{D,\rm{c}}$, $(c)$ $C_{L,\rm{w}}$, and $(d)$ $C_{D,\rm{w}}$ on the streamwise distance $\Delta x$.
  The subscripts c and w denote the cylinder and the airfoil, respectively.
  The cases for $(\alpha,\Delta y)=(30^\circ,0)$ with $\Delta x = [2.0,5.0]$ are shown.  
  }
\label{fig10}
\end{figure}
%%%%%%%%%%%%%%%%%%%%%%%%%%%%%%%%%%%%%%%%

The streamwise distance of the airfoil to the cylinder $\Delta x$ intuitively affects the interaction dynamics. 
Here, let us consider seven streamwise distances of $\Delta x \in [2.0,5.0]$ at a fixed angle of attack $\alpha=30^\circ$ with a vertical position of the airfoil $\Delta y=0$. 
The time series of vorticity fields are presented in figure~\ref{fig9}. 
When the distance between the cylinder and the airfoil is smaller than $\Delta x = 3.0$, {\it weak shedding} and {\em suppression} are observed due to their strong interactions, analogous to the cases for $\alpha \leq 30^\circ$ in figure~\ref{fig5}.

For a larger distance than $\Delta x =3$, the von Kármán vortex shedding suddenly appears from both bodies. 
Its dynamics change markedly toward being more dynamic, based on the time-averaged aerodynamic coefficients in figure~\ref{fig10}.
The drag and lift coefficients of the cylinder are not monotonically increasing but exhibit convex trends, particularly for the drag coefficient as $\alpha$ increases. 
In other words, the cylinder shedding is optimally suppressed with $\Delta x = 3$, while their interactions are decoupled by distancing a wing from the cylinder.
The flow patterns for $\Delta x \geq 3.5$ are basically similar over the phase since the airfoil shedding is primarily due to the upstream shedding at this angle of attack of $\alpha = 30^\circ$.

It is also worth mentioning that the time-averaged drag coefficient of the airfoil $C_{D,\rm{w}}$ shows a negative value for $\Delta x \leq 3$ in figure~\ref{fig10}$(d)$, generating thrust on average. 
This coincides with a previous study~\citep{Suehiro2005} examining the body-body interaction between two cylinders in a tandem arrangement at $Re_D=100$, reporting that the streamwise force in the downstream object is not a drag but rather an impulsive force.

\subsubsection{The influence of vertical distance \texorpdfstring{$\Delta y$}{Delta y}}
\label{sec:dy}

%%%%%%%%%%%%%%%%%%%%%%%%%%%%%%%%%%%%%%%%%%%%%%%
\begin{figure}[t]
  \centerline{
  \includegraphics[width=\textwidth]{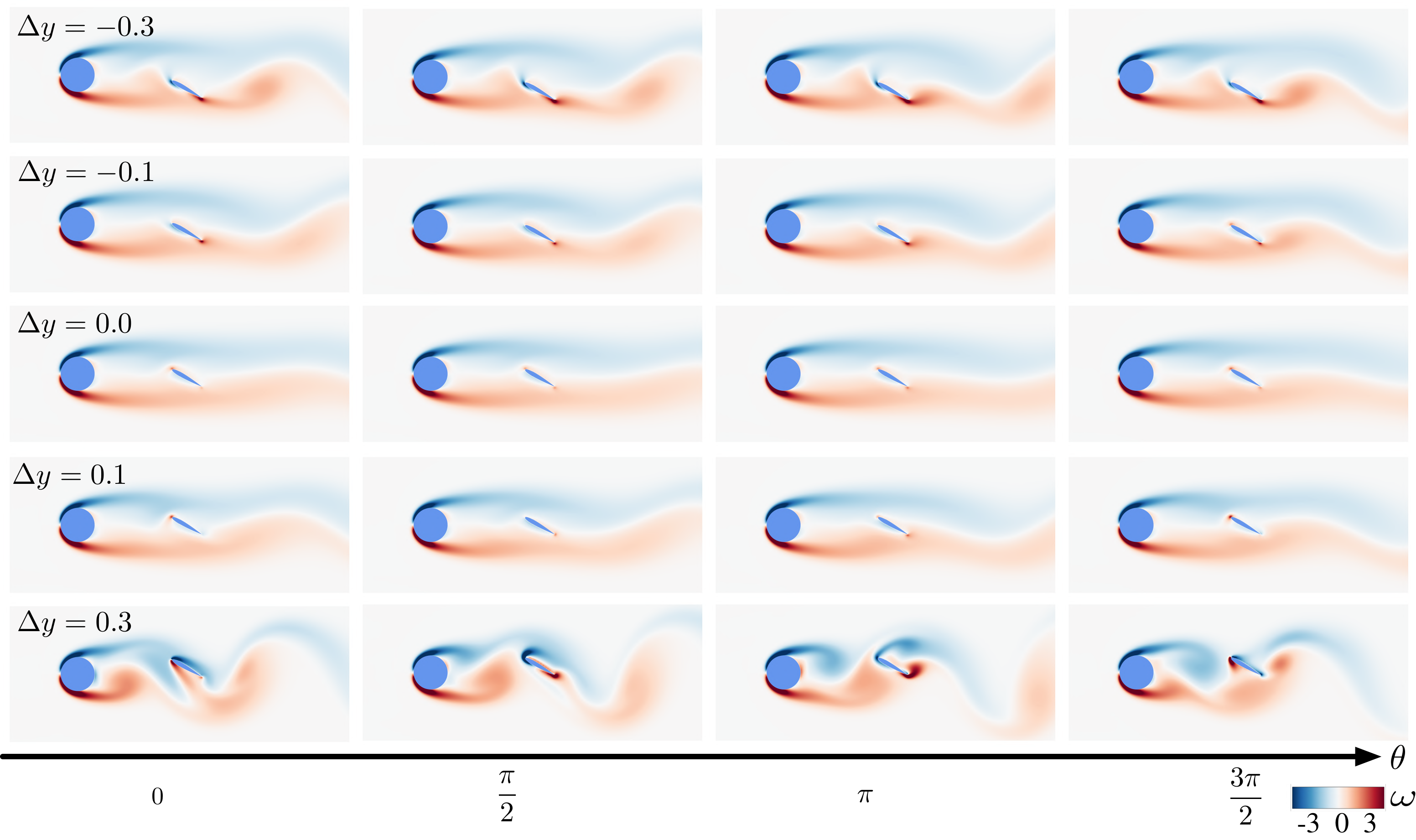}
  }
  \caption{
  The dependence of the vorticity field $\omega$ on the vertical distance $\Delta y$ over the phase $\theta$.
  The phase is defined for each periodic time series of the vorticity field.
  % $\theta \equiv 0$ such that $C'_L (\coloneq C_L -\overline{C}_L )=0$.
  The cases for $(\alpha,\Delta x)=(30^\circ,3.0)$ with $\Delta y = [-0.3,0.3]$ are shown.  
  }
  % \vspace{-5mm}
\label{fig3_10}
\end{figure}
%%%%%%%%%%%%%%%%%%%%%%%%%%%%%%%%%%%%%%%%%%%%%%%

The vertical position of the airfoil relative to the cylinder $\Delta y$ determines how the vortex shedding from the cylinder hits the airfoil, thereby affecting the interaction dynamics.
To examine the dependence of the body-body interaction on the vertical distance, we consider five streamwise distances of $\Delta y \in [-0.3, 0.3]$ for fixed parameters $\alpha=30^\circ$ and $\Delta x=3.0$ as representative examples.

The time series of vorticity fields are presented in figure~\ref{fig3_10}. 
At the current high angle of attack of the airfoil, the leading-edge placement in the cylinder shedding, i.e., with positive $(\Delta y = -0.3)$, near zero ($\Delta y = -0.1, 0, 0.1$), or negative vorticity ($\Delta y=0.3$), can determine the entire shedding patterns.
While {\em shedding with feedback} emerges at $\Delta y=-0.3$, the shedding is suppressed with $\Delta y = -0.1, 0$ and $0.1$ ({\em suppression}).
In contrast, the case at $\Delta y = 0.3$ causes stronger circulations for both bodies, classified into the pattern of {\em shedding from both}.

%%%%%%%%%%%%%%%%%%%%%%%%%%%%%%%%%%%%%%%%%%%%%%%
\begin{figure}
  \centerline{
  \includegraphics[width=0.9\textwidth]{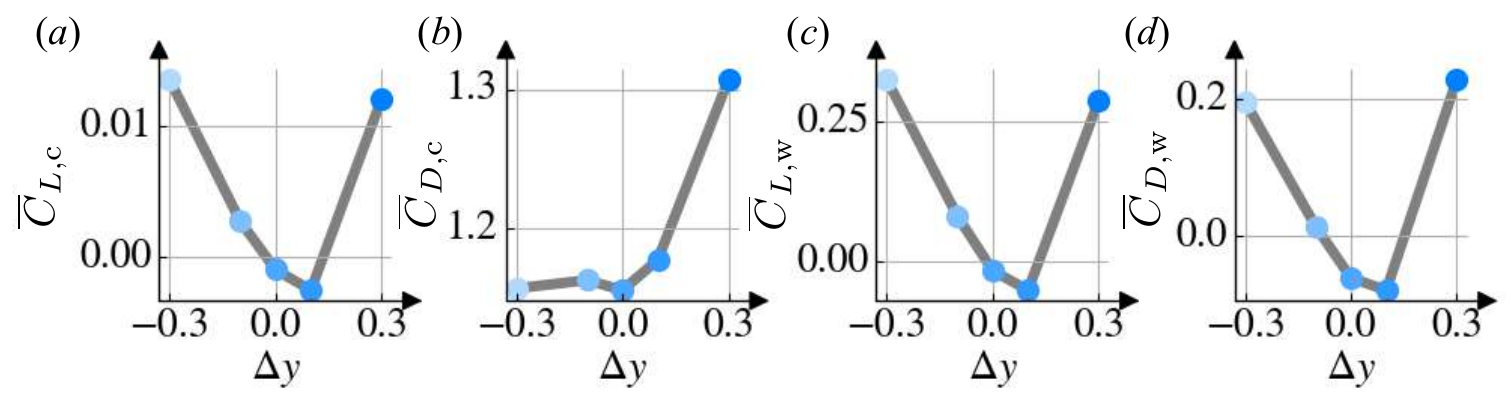}
  }
  % \vspace{-3mm}
  \caption{
  The dependence of aerodynamic coefficients $(a)$ $C_{L,\rm{c}}$, $(b)$ $C_{D,\rm{c}}$, $(c)$ $C_{L,\rm{w}}$, and $(d)$ $C_{D,\rm{w}}$ on the vertical distance $\Delta y$.
  The subscripts c and w denote the cylinder and the airfoil, respectively.
    The cases for $(\alpha,\Delta x)=(30^\circ,3.0)$ with $\Delta y = [-0.3,0.3]$ are shown.    
  }
\label{fig3_11}
\end{figure}
%%%%%%%%%%%%%%%%%%%%%%%%%%%%%%%%%%%%%%%%%%%%%%%

For the flow at $\Delta y = -0.3$, the feedback from the airfoil to the cylinder causes three steady structures: negative vorticity above the leading edge, positive vorticity in front of the leading edge, and on the pressure side of the airfoil.
While these structures between the cylinder and the airfoil almost do not change in time, the unsteadiness and time-varying wake shedding mostly occur from the airfoil.
The time-averaged aerodynamic coefficients of the cylinder and the airfoil, except for the drag coefficient of the cylinder, show larger values than those of other $\Delta y$, as presented in figure~\ref{fig3_11}.
Since the cylinder wake shedding is somewhat suppressed, the drag coefficient of the cylinder is smaller than that at $\Delta y = 0.3$.

Intermediate vertical positions of the airfoil from $\Delta y = -0.1$ to $0.1$ correspond to the region that experiences the slipstream of the cylinder, as shown in figures~\ref{fig3_10} and~\ref{fig3_11}. 
Their vorticity fields show suppressed fluctuations between the two bodies and vortex shedding.
The time-averaged lift and drag coefficients of the airfoil show relatively small values, while the drag coefficient of the cylinder remains almost the same value as for the lower $\Delta y$, corresponding to the suppression of cylinder shedding.

The case with a higher vertical position $\Delta y = 0.3$ shows vortex shedding not only from the airfoil but also from the cylinder, categorized as {\em shedding from both}.
At $\theta = 0$ and $3\pi/2$, the airfoil encounters flow structures with negative vorticity separated from a cylinder, enhancing the separation from the leading edge with the pre-existing negative vorticity near the leading edge.
In addition, the pressure side of the airfoil is greatly affected by the rolled-up positive vorticity from the cylinder shedding for $\pi/2 \leq \theta \leq \pi$, triggering the trailing-edge separation from the airfoil.
These mechanisms can cause coherent shedding structures from both bodies, which uniquely emerge at this particular $\Delta y$ of 0.3.

The vortex shedding from the cylinder dominates the flow around the airfoil at $\Delta y = 0.3$.
All of the lift and the drag coefficients of both bodies exhibit larger values than other $\Delta y$ or are almost the same as the largest value of all, as presented in figure~\ref{fig3_11}. 
This suggests that there is a sweet spot to achieve small drag or lift of the body in the wake appearing approximately just behind the cylinder.

\subsection{Synchronization analysis}

The analysis above shows that there exists an optimal location of the airfoil to suppress the unsteadiness of the current cylinder-airfoil interaction flows.
To further examine the present suppression phenomena, this section studies the flow from the frequency and the synchronization perspectives with the assistance of data-driven techniques.

\subsubsection{Frequency-based analysis}
\label{sec:OD_freq}

%%%%%%%%%%%%%%%%%%%%%%%%%%%%%%%%%%%%%%%%
\begin{figure}[t]
  \centerline{
  \includegraphics[width=1.15\textwidth]{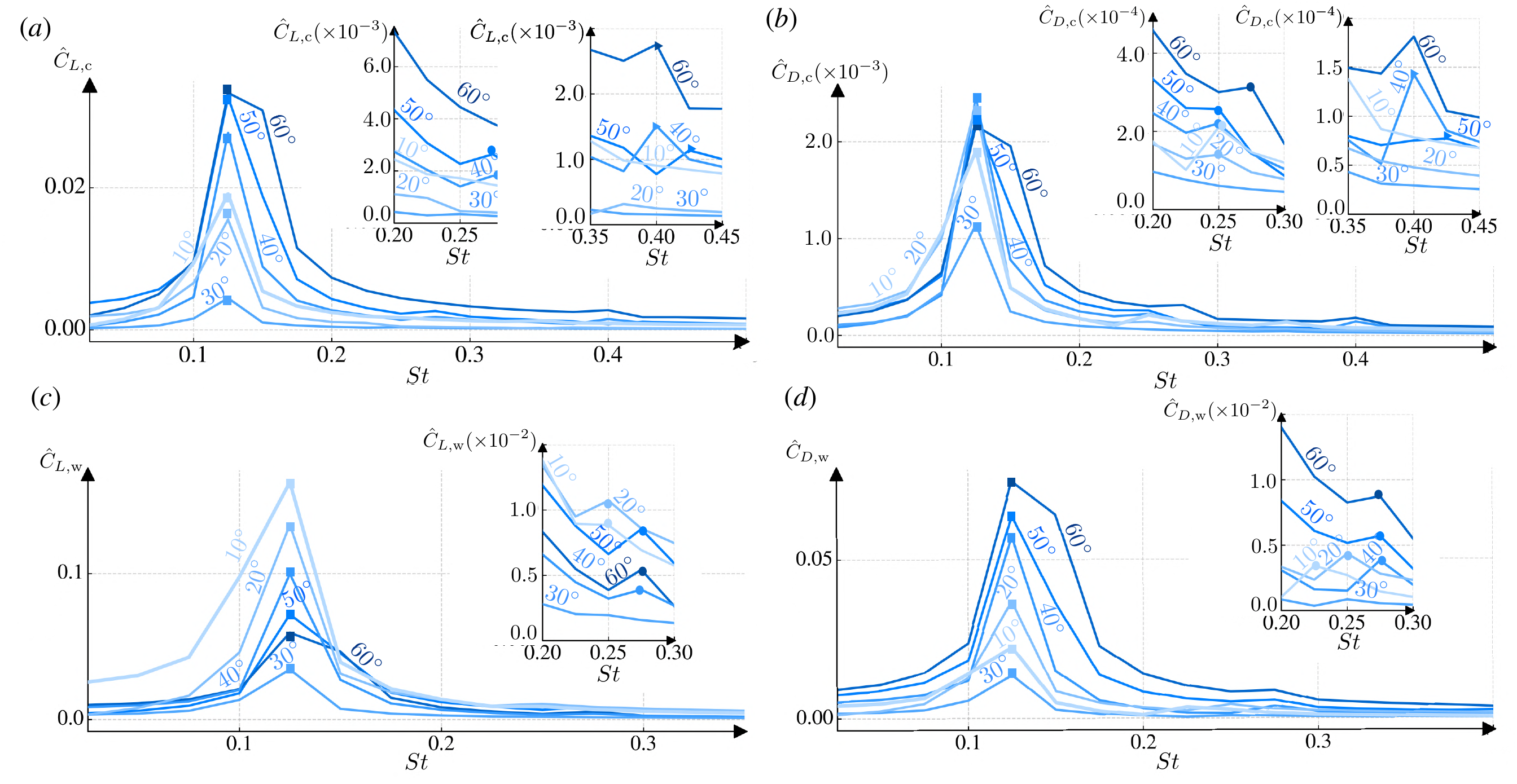}
  }
  \caption{
    Frequency spectra of $(a)$ $C_{L,\rm{c}}$, $(b)$ $C_{L,\rm{w}}$, $(c)$ $C_{D,\rm{c}}$, and $(d)$ $C_{D,\rm{w}}$ for a range of $\alpha$; $\blacksquare$, first dominant frequency; $\bullet$, second dominant frequency; $\blacktriangle$, third dominant frequency. 
    The cases for $(\Delta x,\Delta y)=(3,0)$ with $\alpha = [10,60]^\circ$ are shown.
    }
\label{fig7}
\end{figure}
%%%%%%%%%%%%%%%%%%%%%%%%%%%%%%%%%%%%%%%%

%%%%%%%%%%%%%%%%%%%%%%%%%%%%%%%%%%%%%%%%
\begin{figure}
  \centerline{
  \includegraphics[width=\textwidth]{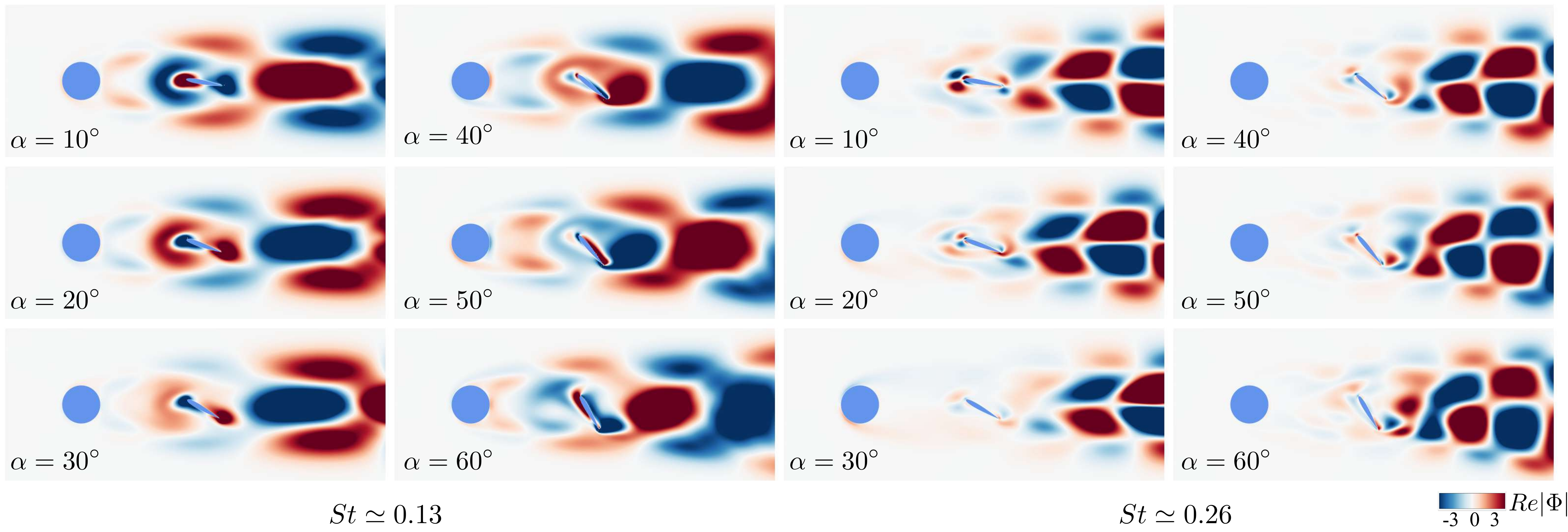}
  }
  \caption{
    Modal structures extracted by dynamic mode decomposition for a range of $\alpha$; each Strouhal number $St$ corresponds to the first and second dominant frequencies shown in figure~\ref{fig5}. 
    The cases for $(\Delta x,\Delta y)=(3,0)$ with $\alpha = [10,60]^\circ$ are shown.
    }
\label{fig8}
\end{figure}
%%%%%%%%%%%%%%%%%%%%%%%%%%%%%%%%%%%%%%%%

As described in section~\ref{sec:method}, this study aims to identify the suppression phenomena on a low-order coordinate that enables performing the linear stability analysis.
Hence, it is anticipated that a comprehensive understanding of flow in the frequency domain would be helpful for the identification of shedding suppression.
Here, we perform the fast Fourier transform for aerodynamic coefficients and the dynamic mode decomposition~\cite[DMD;][]{SCHMID_2010} for vorticity snapshots.
We hereafter consider flows with $(\Delta x,\Delta y)=(3,0)$ for $\alpha = [10,60]^\circ$, discussed in figures~\ref{fig5} and \ref{fig6}, as representative cases.

The frequency spectra of the lift and drag history of both the cylinder and the airfoil for the selected cases are depicted in figure~\ref{fig7}. 
A sampling time sufficient to capture seven dominant shedding cycles is used to perform the spectral analysis.
The first dominant frequency at $St\simeq0.13$ captured for all aerodynamic coefficients corresponds to the von Kármán vortex shedding from the airfoil. 
However, the order of lift magnitude for the current interaction flow is not in the order of angle of attack, unlike the single airfoil flow whose unsteadiness solely increases with the angle of attack at this Reynolds number~\citep{Maji2024,Fukami2024_attenuation}.
The magnitude of the first dominant frequency is suppressed at $\alpha=30^\circ$ due to the interaction, consistent with observations in the flow field.

These observations can be analyzed with the DMD modes on the left side of figure~\ref{fig8}. 
Bluff structures around the leading edge are observed at $St\simeq0.13$ in the cases of $\alpha=10^\circ$ and $20^\circ$.
Their mode intensities and structural sizes become small for $\alpha \geq 30^\circ$.
There is also another prominent modal structure right behind the cylinder.
Interestingly, this structure is weakened only in the case of $\alpha = 30^\circ$.
This indicates that the balance between the bluff structure around the leading edge of the downstream airfoil and the modal structure inbetween the bodies at $St\simeq0.13$ can determine the shedding suppression observed in this study.

In addition, the second dominant frequencies are observed at $St\simeq0.26$, as depicted in figure~\ref{fig7}. 
Circle markers on the zoomed-in view in figure~\ref{fig7} correspond to the second dominant frequency of aerodynamic coefficients.
The frequency spectra of the airfoil aerodynamic coefficients, presented in figures~\ref{fig7}$(c)$ and $(d)$, show that the secondary peak for $\alpha=10^\circ$ and $20^\circ$ exists in a different frequency band from the higher $\alpha$.
Furthermore, the second peak disappears for $\alpha = 30^\circ$.

The observation above is comparable with the DMD modes for the second dominant frequency presented in the right-side portion of figure~\ref{fig8}.
Two small-sized structures appear around the leading edge of the airfoil, which become weaker as $\alpha$ increases. 
This structure almost disappears at $\alpha=30^\circ$, corresponding to {\em suppression} of the fluctuation between the cylinder and the airfoil. 
For higher $\alpha$, instead of the structures around the leading edge, a triangular structure near the trailing edge emerges, and its size becomes larger as $\alpha$ increases, corresponding to the magnitude of the second dominant frequencies.
This suggests that the balance between the small-sized structures near the leading edge and the triangular structure near the trailing edge at $St\simeq0.26$ may also contribute to the current suppression phenomena.

\subsubsection{Oscillation death}
\label{sec:OD_od}

Based on the above analysis in the frequency domain, we have found that the flow field with an airfoil at $\alpha = 30^\circ$ presents intrinsic characteristics compared to other cases for $(\Delta x,\Delta y)=(3,0)$.
This subsection aims to identify the suppression phenomena emerging at $\alpha = 30^\circ$ from aerodynamic responses with data-driven techniques.
This is achieved by using a synchronization-theoretic autoencoder while relating the current suppressed wake shedding to oscillation death.
As described in section~\ref{Chapter2_2_1}, this enables the projection of aerodynamic coefficients for each of the cylinder and the airfoil into a low-order subspace that can be modeled as two coupled Stuart--Landau oscillators~\citep{Stuart1958, Landau1944,araya2026inferring}.

%%%%%%%%%%%%%%%%%%%%%%%%%%%%%%%%%%%%%%%%%%%%%
\begin{figure}
  \centerline{
  \includegraphics[width=1\textwidth]{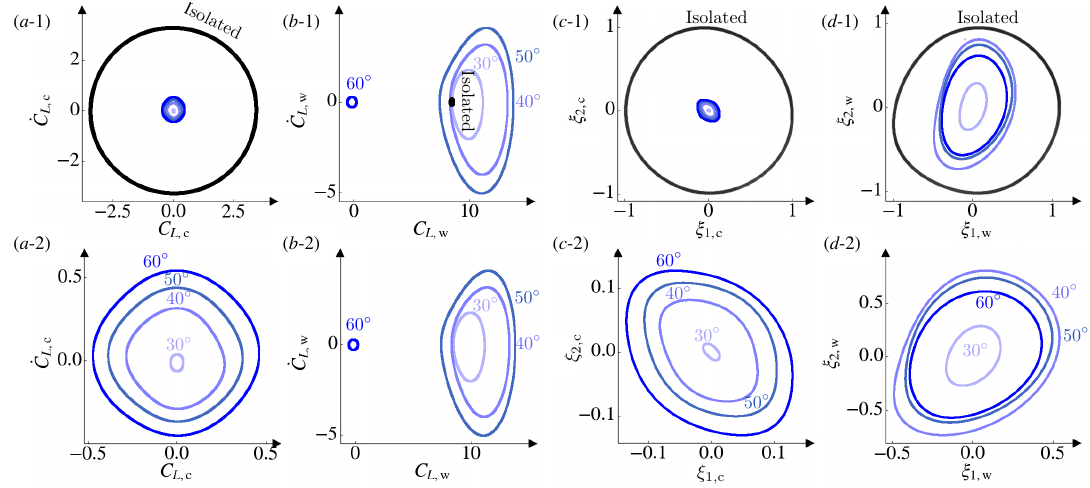}
  }
  % \vspace{-3mm}
  \caption{
   $(a)$ and $(b)$: The lift coefficient and its time derivative for the single-body flows, and the cylinder-airfoil interaction flows with $(\Delta x,\Delta y)=(3,0)$ over $\alpha \in [30^\circ, 60^\circ]$.
   The phase space of ($a$-1) cylinder flow and ($b$-1) airfoil flow.
   Their zoomed-in view for ($a$-2) cylinder flow and ($b$-2) airfoil flow.
   The latent trajectories obtained from the synchronization-theoretic autoencoder of ($c$-1) cylinder flow and ($d$-1) airfoil flow.
   Their zoomed-in view for ($c$-2) cylinder flow and ($d$-2) airfoil flow.
  }
\label{fig_syn1}
\end{figure}
%%%%%%%%%%%%%%%%%%%%%%%%%%%%%%%%%%%%%%%%%%%%%

Let us begin the discussion here in the phase space composed of the lift coefficient $C_L$ and its time derivative $\dot{C_L}$.
% This study chooses the lift coefficient $C_L$ and its time derivative $\dot{C_L}$ as the input of the autoencoder.
This is because the $C_L-\dot{C_L}$ plane captures the periodic limit-cycle oscillations of the current flows, as shown in figure~\ref{fig_syn1}.
To first model the isolated or non-interactive dynamics of each body as a Stuart--Landau oscillator in equation~\ref{eq:integrated_system}, they need to be a pure circle in phase space due to its definition.
However, while the trajectory of the single cylinder presents a circle, the airfoil counterpart is skewed and squished, as presented in figure~\ref{fig_syn1}($a$-1) and ($b${-1}).
This is due to the asymmetricity of wake shedding for the single airfoil case.
In addition, the relative position of the latent trajectory for the single airfoil case against other interaction cases is not well organized compared to that for the single cylinder case, since how the airfoil interacts with the cylinder shedding nonlinearly varies across the angle of attack.
These make the direct application of Stuart--Landau oscillator-based modeling for the aerodynamic coefficients difficult even if the trajectories themselves present the periodicity.

To address this matter, let us apply the synchronization-theoretic autoencoder-based coordinate transformation for $C_L$ and $\dot{C_L}$, as introduced in section~\ref{Chapter2_2_2}.
The lift coefficient and its time derivative for the single cylinder and the airfoil at $\alpha=30^\circ$ are transformed to be unit circles in the latent space, as shown in figures~\ref{fig_syn1}($c$-1) and ($d$-1).
The isolated dynamics for the cylinder $\dot{{\bm \xi}_c}$ and the airfoil $\dot{{\bm \xi}_c}$ are then described with the operation in equation~\ref{eq:isolatedSINDy} such that 
\begin{equation}
\begin{cases}
    \dot{\boldsymbol{\xi}}_{\rm c} = 
    \begin{bmatrix} 1 & 1.31 \\ -1.31 & 1 \end{bmatrix} \boldsymbol{\xi}_{\rm c} 
    - \|\boldsymbol{\xi}_{\rm c}\|^2 
    \begin{bmatrix} 1 & 0.271 \\ -0.271 & 1 \end{bmatrix} \boldsymbol{\xi}_{\rm c} \\[10pt]
    \dot{\boldsymbol{\xi}}_{\rm w} = 
    \begin{bmatrix} 1 & 2.03 \\ -2.03 & 1 \end{bmatrix} \boldsymbol{\xi}_{\rm w} 
    - \|\boldsymbol{\xi}_{\rm w}\|^2 
    \begin{bmatrix} 1 & 0.607 \\ -0.607 & 1 \end{bmatrix} \boldsymbol{\xi}_{\rm w} 
\end{cases}
\label{eq:3_1}
\end{equation}
where the eigenvalues of this equation obtained via linear stability analysis from the origin are $1.00\pm1.31i$ and $1.00\pm2.03i$, respectively.
Their imaginary parts correspond to $d_{\mathrm{c}}$ and $d_{\mathrm{w}}$ in equations~(\ref{eq:integrated_system}).

The latent trajectories of the interaction case for $(x_0,y_0)=(3,0)$ and $\alpha\in[30,60]$ are presented in in figure~\ref{fig_syn1}$(c)$ and $(d)$.
They show the interaction effect between the cylinder and the airfoil as the deformation and its magnitude change from the unit circle, thereby forming the ellipses.
For the cylinder latent trajectories, the interaction magnitude is proportional to the time-averaged lift coefficient, as shown in figure~\ref{fig6}.
In contrast, the latent trajectories for the airfoil do not show an exact correspondence to the lift fluctuation, e.g., as seen from the mapping location for the trajectory at $\alpha = 60^\circ$ in figures~\ref{fig_syn1}$(b)$-2 and $(d)$-2.
We however note that this rather does not need to be consistent with each other, since the shape and position of the ellipse are determined based on the difference from the non-interactive flow.

%%%%%%%%%%%%%%%%%%%%%%%%%%%%%%%%%%%%%%%%%%%%%
\begin{figure}
  \centerline{
  \includegraphics[width=1.1\textwidth]{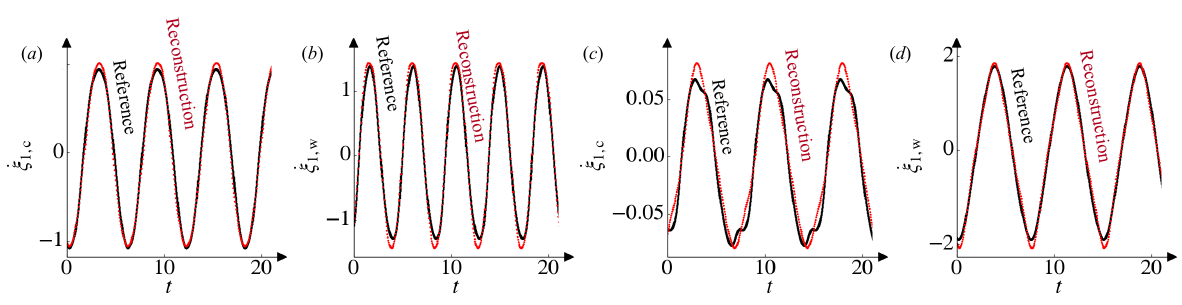}
  }
  % \vspace{-3mm}
  \caption{
  Comparison between the reference and reconstructed latent variables for $(a)$ the isolated cylinder, $(b)$ the isolated airfoil, $(c)$ the interactive cylinder, and $(d)$ the interactive airfoil for the case of $(\Delta x, \Delta y, \alpha)=(3,0, 30^\circ)$.
  }
\label{fig_syn2}
\end{figure}
%%%%%%%%%%%%%%%%%%%%%%%%%%%%%%%%%%%%%%%%%%%%%

The reconstructed latent variables using the identified model are compared to the reference trajectories, as shown in figure~\ref{fig_syn2}.
Here, the flow with $(\Delta x, \Delta y, \alpha)=(3,0, 30^\circ)$ is considered as a representative case of cylinder-airfoil interaction.
They are in agreement with each other, implying that the current cylinder-airfoil interactions can indeed be modeled as the two coupled Stuart--Landau oscillators in the autoencoder-based subspace.

%%%%%%%%%%%%%%%%%%%%%%%%%%%%%%%%%%%%%%%%%%%%%
\begin{figure}
  \centerline{
  \includegraphics[width=0.95\textwidth]{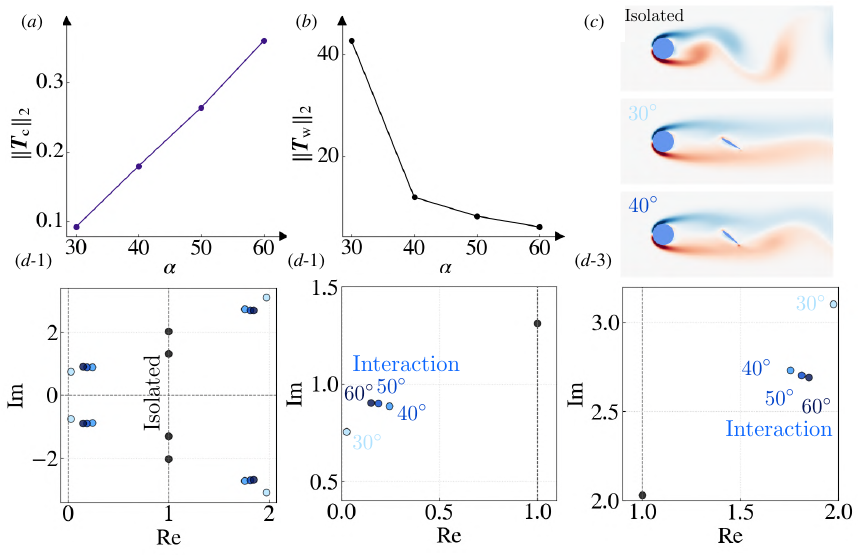}
  }
  % \vspace{-3mm}
  \caption{
    $(a)$ and $(b)$ The dependence of the $L_2$ norm of the interaction matrix for the cylinder and the airfoil, respectively, on the angle of attack.
    $(c)$ The flow fields corresponding to the eigenvalue.
    ($d$-1) The eigenvalues on the Gaussian plane. Stronger color corresponds to a higher angle of attack. Black dots are for the isolated cases. ($d$-2) and ($d$-3) expanded plots of the plot ($d$-1).
  }
\label{fig_syn3}
\end{figure}
%%%%%%%%%%%%%%%%%%%%%%%%%%%%%%%%%%%%%%%%%%%%%

Based on the identified model, let us examine the interaction coefficients $\boldsymbol{T}_{\mathrm{c}}$ and $\boldsymbol{T}_{\mathrm{w}}$ in the right-hand side of equations~(\ref{eq:integrated_system}), as summarized in figure~\ref{fig_syn3}.
The interaction magnitudes are shown in figures~\ref{fig_syn3}$(a)$ and $(b)$, computed by taking the $L_2$ norm of the matrices $\boldsymbol{T}_{\mathrm{c}}$ and $\boldsymbol{T}_{\mathrm{w}}$ containing the interaction coefficients.
As discussed in section~\ref{sec:alpha}, the feedback from the airfoil becomes stronger as $\alpha$ increases. 
This trend of the feedback from the airfoil to the cylinder is reflected in the increasing trend in figure~\ref{fig_syn3}$(a)$. 
For the norm of $\boldsymbol{T}_{\mathrm{w}}$ that represents the interaction from the cylinder to the airfoil, the value at $\alpha = 30$ is much greater than those at higher $\alpha$, corresponding to {\em suppression} by the interaction depicted in figure~\ref{fig_syn3}$(c)$.
The value of $\| \boldsymbol{T}_{\mathrm{w}}\|_2$ then decreases as $\alpha$ increases, corresponding to the fact that the influence of the angle of attack of the airfoil dominates with the increment of $\alpha$.
This coincides with the trend found in figure~\ref{fig5}; at larger $\alpha$, the cylinder and the airfoil tend to act as isolated bodies.

The eigenvalues obtained from linear stability analysis for the identified coupled Stuart--Landau oscillator models are shown in figure~\ref{fig_syn3}$(d)$ on the Gaussian plane.
The eigenvalue for $\alpha=30^\circ$ shows ``oscillation death", as evident from its near-zero real value.
It is worth pointing out that the real part of the identified eigenvalue becomes smaller and heads toward zero for $\alpha \geq 40^\circ$, despite the fact that it originally increases for $30^\circ \leq \alpha \leq 40^\circ$.
This suggests that the interaction flow at $\alpha = 40^\circ$ may be regarded as a transitional state between the suppression flow at $\alpha = 30^\circ$ and the shedding-with-feedback flow at $\alpha = 60^\circ$.
Another eigenvalue for $\alpha=30^\circ$ shows stronger instability than at higher angles of attack.
This suggests the existence of other solutions at $\alpha=30^\circ$, coinciding with the observation in Mizushima $\&$ Suehiro (2005) \citep{Suehiro2005}.

\section{Concluding remarks}
\label{sec:conc}

We examined the flow interaction between a cylinder and a NACA0012 airfoil using direct numerical simulation and machine learning. 
We considered a two-dimensional incompressible flow at a Reynolds number $Re = 100$, where the airfoil is positioned in the wake of a cylinder. 
By sweeping the parameter space, composed of the distance between the cylinder and the airfoil $\Delta x$, the relative height $\Delta y$, and the angle of attack of the airfoil $\alpha$, four characteristic vortex-shedding regimes were observed: {\em weak shedding}, {\em suppression}, {\em shedding with feedback}, and {\em shedding from both}.
The effects of each parameter were examined with representative flows.
The streamwise distance between the cylinder and the airfoil $\Delta x$ affects the intensity of the interaction between the two bodies. The relative height $\Delta y$ shows how the two bodies interact with each other. The angle of attack of the airfoil $\alpha$ influences the flow instability around the airfoil and the feedback magnitude from the pressure side of the airfoil to the cylinder.

We particularly focused on the suppressed vortex shedding at a certain arrangement as oscillation death, with the theoretical prediction of flow stability on a machine-learned low-order coordinate. 
Independent flows around each body were regarded as a system that exhibits isolated periodic vortex shedding at infinite distance, and interactions appeared at smaller distances.
We proposed a synchronization-theoretic autoencoder to derive models for two coupled Stuart--Landau oscillators from an aerodynamic coefficient.
The current autoencoder enables the extraction of the coupling magnitude for high-dimensional vortex interactions in an interpretable form while representing them as the deformation from the pure circle of the isolated body dynamics to an ellipse.
Their stability was further examined through linear stability analysis in the latent space, capturing the oscillation death of the current cylinder-airfoil interaction flow in a low-order manner.

While this study focused on the interaction between the lift coefficients of two bodies, the present synchronization-theoretic approach can take other variables, such as the drag coefficient and lift-drag ratio, if they capture their flow characteristics well in a high-dimensional space.
As choosing such a variable is generally not trivial, one can consider performing the autoencoder-based compression for flow snapshots in the future~\citep{FF2025,Fukami_JFMPerspective}.
This flexibility highlights the potential of the present technique to provide stability-based insights into a range of unsteady flow interaction problems.

\section*{Acknowledgments}
{
{
Kai Fukami acknowledges support from the JSPS KAKENHI Grant No. JP25K23418 and No. JP26K01129, the JST PRESTO Grant No. JPMJPR25KA, and the MEXT Coordination Funds for Promoting Aerospace Utilization Grant No. JPJ000959.
Yuzuru Kato acknowledges support from the JST PRESTO Grant No. JPMJPR24K3.
}}

% \vspace{-5mm}
\section*{Declaration of interests}
{The authors report no conflict of interest.}

%%%%%%%%%%%%%%%%%%%%%%%%%%%%%%%%
%%%%%%%%%%%%%%%%%%%%%%%%%%%%%%%%%
%%%%%%%%%%%%%%%%%%%%%%%%%%%%%%%%%

\bibliographystyle{unsrt}  
\bibliography{refs}
%%% Remove comment to use the external .bib file (using bibtex).

\end{document}